\documentclass[aps, prx, twocolumn, superscriptaddress, longbibliography, floatfix]{revtex4-1}
\usepackage{natbib}
\usepackage{amsmath} 
\usepackage{amssymb}
\usepackage{graphicx}
\usepackage{times}
\usepackage{wasysym}
\usepackage{hyperref}

\begin{document}
\newcommand\bbone{\ensuremath{\mathbbm{1}}}
\newcommand{\ul}{\underline}
\newcommand{\bp}{{\bf p}}
\newcommand{\vl}{v_{_L}}
\newcommand{\vc}{\mathbf}
\newcommand{\be}{\begin{equation}}
\newcommand{\ee}{\end{equation}}
\newcommand{\bk}{{{\bf{k}}}}
\newcommand{\bK}{{{\bf{K}}}}
\newcommand{\cE}{{{\cal E}}}
\newcommand{\bQ}{{{\bf{Q}}}}
\newcommand{\br}{{{\bf{r}}}}
\newcommand{\bg}{{{\bf{g}}}}
\newcommand{\bG}{{{\bf{G}}}}
\newcommand{\hbr}{{\hat{\bf{r}}}}
\newcommand{\bR}{{{\bf{R}}}}
\newcommand{\bq}{{\bf{q}}}
\newcommand{\hx}{{\hat{x}}}
\newcommand{\hy}{{\hat{y}}}
\newcommand{\hd}{{\hat{\delta}}}
\newcommand{\bea}{\begin{eqnarray}}
\newcommand{\eea}{\end{eqnarray}}
\newcommand{\ra}{\rangle}
\newcommand{\la}{\langle}
\renewcommand{\tt}{{\tilde{t}}}
\newcommand{\upa}{\uparrow}
\newcommand{\dna}{\downarrow}
\newcommand{\bS}{{\bf S}}
\newcommand{\vS}{\vec{S}}
\newcommand{\dg}{{\dagger}}
\newcommand{\pdg}{{\phantom\dagger}}
\newcommand{\tphi}{{\tilde\phi}}
\newcommand{\cf}{{\cal F}}
\newcommand{\ca}{{\cal A}}
\renewcommand{\ni}{\noindent}
\newcommand{\ct}{{\cal T}}
\newcommand{\brf}{\bar{F}}
\newcommand{\brg}{\bar{G}}
\newcommand{\jeff}{j_{\rm eff}}

\newcommand{\BaCeIrO}{Ba$_{\rm 2}$CeIrO$_{\rm 6}$}
\newcommand{\NaIrO}{Na$_{\rm 2}$IrO$_{\rm 3}$}
\newcommand{\LiIrO}{Li$_{\rm 2}$IrO$_{\rm 3}$}
\newcommand{\HLiIrO}{H$_{\rm 3}$LiIr$_{\rm 2}$O$_{\rm 6}$}
\newcommand{\RuCl}{$\alpha$-RuCl$_{\rm 3}$}
\newcommand{\SrIrO}{Sr$_{\rm 2}$IrO$_{\rm 4}$}

\title{Spin-orbit entangled ${\bf j=1/2}$ moments in \BaCeIrO\ -- a  frustrated \textit{fcc} quantum magnet}

\author{A. Revelli}
\affiliation{Institute of Physics II, University of Cologne, 50937 Cologne, Germany}
\author{C.C. Loo}
\affiliation{Institute of Physics II, University of Cologne, 50937 Cologne, Germany}
\author{D. Kiese}
\affiliation{Institute for Theoretical Physics, University of Cologne, 50937 Cologne, Germany}
\author{P. Becker}
\affiliation{Sect. Crystallography, Institute of Geology and Mineralogy, University of Cologne, 50674 Cologne, Germany}
\author{T. Fr\"ohlich}
\affiliation{Institute of Physics II, University of Cologne, 50937 Cologne, Germany}
\author{T. Lorenz}
\affiliation{Institute of Physics II, University of Cologne, 50937 Cologne, Germany}
\author{M. Moretti Sala}
\affiliation{Dipartimento di Fisica, Politecnico di Milano, I-20133 Milano, Italy}
\author{G.~Monaco}
\affiliation{Dipartimento di Fisica, Universit\`{a} di Trento, I-38123 Povo (TN), Italy}
\author{F.L. Buessen}
\affiliation{Institute for Theoretical Physics, University of Cologne, 50937 Cologne, Germany}
\author{J. Attig}
\affiliation{Institute for Theoretical Physics, University of Cologne, 50937 Cologne, Germany}
\author{M. Hermanns}
\affiliation{Department of Physics, Stockholm University, AlbaNova University Center, SE-106 91 Stockholm, Sweden}
\affiliation{Nordita, KTH Royal Institute of Technology and Stockholm University, SE-106 91 Stockholm, Sweden}
\author{S.V.~Streltsov}
\affiliation{\mbox{M.N.\ Mikheev Institute of Metal Physics, Ural Branch, Russian Academy of Sciences, 
		620137 Ekaterinburg, Russia}}
\affiliation{Ural Federal University, 620002 Ekaterinburg, Russia}
\author{D.I. Khomskii}
\affiliation{Institute of Physics II, University of Cologne, 50937 Cologne, Germany}
\author{J. van den Brink}
\affiliation{Institute for Theoretical Solid State Physics, IFW Dresden, 01069 Dresden, Germany}
\author{M. Braden}
\affiliation{Institute of Physics II, University of Cologne, 50937 Cologne, Germany}
\author{P.H.M. van Loosdrecht}
\affiliation{Institute of Physics II, University of Cologne, 50937 Cologne, Germany}
\author{S. Trebst}
\affiliation{Institute for Theoretical Physics, University of Cologne, 50937 Cologne, Germany}
\author{A. Paramekanti}
\affiliation{Department of Physics, University of Toronto, Toronto, Ontario M5S 1A7, Canada}
\author{M. Gr\"uninger}
\affiliation{Institute of Physics II, University of Cologne, 50937 Cologne, Germany}

\begin{abstract}
We establish the double perovskite \BaCeIrO\ as a nearly ideal model system for $j$\,=\,1/2 moments, with 
resonant inelastic x-ray scattering indicating that the ideal $j$\,=\,1/2 state contributes by more 
than 99\,\% to the ground-state wavefunction.
The local $j$\,=\,1/2 moments form an {\em fcc} lattice and are found to order antiferromagnetically at 
$T_N$\,=\,14\,K, more than an order of magnitude below the Curie-Weiss temperature. 
Model calculations show that the geometric frustration of the {\em fcc} Heisenberg antiferromagnet is 
further enhanced by a next-nearest neighbor exchange, and a significant size of the latter is 
indicated by {\em ab initio} theory. Our theoretical analysis shows that magnetic 
order is driven by a bond-directional Kitaev exchange and by local distortions via a strong magneto-elastic effect. 
Both, the suppression of frustration by Kitaev exchange and the strong magneto-elastic effect
are typically not expected for $j$\,=\,$1/2$ compounds making \BaCeIrO\ a riveting example 
for the rich physics of spin-orbit entangled Mott insulators.
\end{abstract}

\date{January 17, 2019; revised version: July 21, 2019}

\maketitle

\section{Introduction}

Spin-orbit entangled Mott insulators stand out in the growing family of quantum materials with strong 
spin-orbit coupling for their correlation-driven phenomena \cite{WitczakKrempa2014}. 
Of particular interest are materials with partially filled $4d$ and $5d$ orbitals, such as the iridates,
in which the formation of local $j$\,=\,$1/2$ moments is an iridescent source of rich physics \cite{Rau2016}.
The spin-orbit entangled wavefunction of these Kramers doublets gives rise to fundamentally different 
types of exchange interactions depending on the geometric arrangement of the elementary octahedral
IrO$_6$ building blocks \cite{Khaliullin2005,Chen2008,Jackeli2009}.
Corner-sharing octahedra yield isotropic Heisenberg exchange, which has been explored 
as a potential source of spin-orbit assisted superconductivity
\cite{Kim2014,Torre2015,Cao2016a,Yan2015,Kim2016d}
in the context of Sr$_{\rm 2}$IrO$_{\rm 4}$ \cite{Kim2008,Kim2009}, 
an isostructural analogue of the high-T$_c$ parent compound La$_{\rm 2}$CuO$_{\rm 4}$.
Edge-sharing octahedra, in contrast, give rise to Kitaev-type bond-directional exchange, 
which has initiated an intense search for spin-orbit driven frustrated quantum magnetism
in so-called Kitaev materials \cite{Trebst2017} such as the honeycomb iridates \NaIrO, $\alpha$-\LiIrO, 
and \HLiIrO\ \cite{Singh2010,Singh2012,Kitagawa2018} and the related \RuCl\ \cite{Plumb2014}.
Possibly the most spectacular experimental result in this realm is the recent claim 
of a quantized thermal Hall effect in \RuCl\ \cite{Kasahara2018}, a direct signature of the long sought-after 
Kitaev spin liquid \cite{Kitaev2006}.

\begin{figure}[b]
	\centering
	\includegraphics[width=.96\columnwidth]{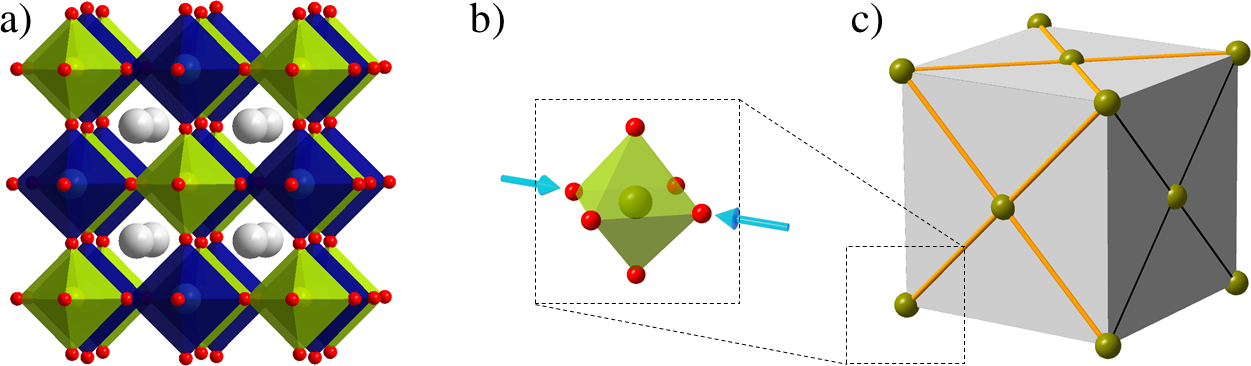}
	\caption{a) {\bf Cubic double-perovskite structure} as observed in x-ray diffraction.
		Small green (large blue) octahedra are centered around 
	the Ir$^{4+}$ (Ce$^{4+}$) sites. Light gray (red) spheres depict Ba$^{2+}$ (O$^{2-}$) ions. 
	Each Ir moment is coupled to 12 nearest neighbors (orange and black lines in c)) 
	and to 6 next-nearest neighbors (along the edges of the cube in c)), enhancing the frustration. 	
	Our RIXS data reveal local distortions from cubic symmetry. Assuming that the 
	distortion is tetragonal, as schematically illustrated in b), we find  
    a massive bond-dependent variation of the nearest-neighbor exchange constants between Ir moments 
    as depicted in c), where couplings indicated in orange and black have different strength, 
    reducing the frustration. 
\label{fig:TetragonalDistortion}
	}
\end{figure}

In this paper, we first demonstrate experimentally that the double perovskite \BaCeIrO\ 
is a nearly ideal realization of a $j$\,=\,$1/2$ Mott insulator, forming a model system for frustrated 
quantum magnetism on the \textit{fcc} lattice. Our x-ray diffraction results show a global cubic $Fm\bar{3}m$ 
structure, while resonant inelastic x-ray scattering (RIXS) reveals a small non-cubic distortion resulting 
in a ground-state wavefunction which overlaps by more than 99\,\% with the ideal cubic $j$\,=\,1/2 state. 
The magnetic susceptibility shows an antiferromagnetic ordering temperature $T_N$\,=\,14\,K which is 
suppressed by more than an order of magnitude compared to the  Curie-Weiss temperature $|\Theta_{\rm CW}|$, 
resulting in a large frustration parameter $f$\,=\,$|\Theta_{\rm CW}|/T_N \gtrsim 13$.
Employing a combination of density functional theory and microscopic model simulations, 
we address the minimal model for \BaCeIrO\ and its phase diagram. 
The system shows a particularly high degree of frustration, since the geometric frustration of antiferromagnetic 
nearest-neighbor Heisenberg exchange on the \textit{fcc} lattice is augmented by 
next-nearest-neighbor Heisenberg coupling, yielding a wide window of a quantum spin liquid ground state. 
However, an antiferromagnetic
Kitaev-type bond-directional exchange is found to \textit{counteract} this geometric frustration 
and turns out to be instrumental in stabilizing long-range magnetic order -- in contrast to the common wisdom 
that Kitaev interactions in $j$\,=\,$1/2$ compounds enhance frustration and induce spin liquid physics.

It is also common wisdom that the
$j$\,=\,$1/2$ wavefunction does not show orbital degeneracy and hence is not Jahn-Teller active. 
Commonly, this is interpreted as a protection of $j$\,=\,$1/2$ physics 
against lattice distortions, and small deviations from cubic symmetry with the concomitant change 
of the wavefunction are typically neglected. 
We challenge this point of view and provide theoretical evidence for a strong magneto-elastic effect. 
Our theoretical analysis shows that
even small deviations from the $j$\,=\,$1/2$ wavefunction, associated with small lattice distortions, 
yield a massive bond-dependent variation of the nearest-neighbor 
exchange constants, as illustrated in Fig.\ \ref{fig:TetragonalDistortion}c), effectively 
lifting the strong magnetic frustration.
This dramatic magneto-elastic coupling is of general importance in the quest for exotic spin liquids 
based on $j$\,=\,$1/2$ compounds.

\section{Synthesis and structure}

Single crystals of \BaCeIrO\ of about 1\,mm$^3$ size were grown by melt solution growth (see Appendix A).
X-ray diffraction shows a well ordered double perovskite with Ce-Ir order as 
illustrated in Fig.\ \ref{fig:TetragonalDistortion}a. The cation order
can be explained by the notably different bond lengths of 2.20\,\AA{} for Ce-O and 2.04\,\AA{} for Ir-O.\@ 
For $5d^5$ Ir$^{4+}$, the formation of ideal $j$\,=\,$1/2$ moments requires a cubic crystal field. 
Thus far, deviations from cubic symmetry were reported for all $5d^5$ iridate compounds \cite{Rau2016,Trebst2017}, 
as discussed in more detail in the section on RIXS below. 

For \BaCeIrO, our powder diffraction peaks -- measured using a Stoe Stadi MP 198 powder diffractometer --
are very well described in the cubic space group $Fm\bar{3}m$ 
with a lattice constant of 8.47\AA{} at 300\,K.\@ However, we find a clear broadening of Bragg peaks 
in particular for large diffraction angles $2\theta > 80^\circ$. 
Such broadened Bragg peaks may explain a previous claim of tiny ($< \! 0.2\,\%$) monoclinic distortions 
of the metric in polycrystalline \BaCeIrO\ \cite{Wakeshima2000}. 
Note that the issue of cubic or non-cubic symmetry is often discussed controversially in double perovskites, 
for instance for the closely related Ba$_2$PrIrO$_6$ \cite{Wakeshima2000,Kockelmann06}. 

To resolve this issue, we collected single-crystal x-ray diffraction data. 
Our results strongly support a cubic structure of \BaCeIrO. 
	We employed a Bruker X8 Apex diffractometer, a sample with octahedral shape ($\{111\}$ faces), 
	and a distance to the center of 12.5\,$\mu$m. 
	At room temperature (100 K) 32921 (13274) Bragg reflection intensities were recorded, 
	yielding 198 (205) independent reflections in space group $Fm\bar{3}m$. 
	The single-crystal data do not yield any evidence for significant superstructure 
	reflections with respect to $Fm\bar{3}m$, 
neither at room temperature nor at 100\,K  (see Appendix A).
From this and the description of the powder diffraction pattern with the cubic lattice 
we must conclude that the average structure of Ba$_2$CeIrO$_6$ is cubic.

\begin{table}[b]
	\caption{Results of structure refinements with single-crystal X-ray data obtained 
	at room temperature and 100\,K.\@ 
	In space group $Fm\bar{3}m$ Ba occupies an $8c$ site at (1/4,1/4,1/4), Ce a $4a$ site at (0,0,0), 
	Ir a $4b$ site at (0,0,1/2), and O a $24e$ site at ($x$,0,0). 
	Thermal parameters are given in $10^{-5}$\AA$^2$, and only the O parameters are anisotropic. 
	The weighted $R$ values for the structure factor
    amount to 4.22\,\% and 3.96\,\% at 300\,K and 100\,K, respectively, and refining occupations 
    reduces them to 2.57\,\% and 3.11\,\%, respectively. 
	The occupation of Ba and O was fixed. For the occupation of Ce and Ir we find $99.6$\% ($99.1$\%) 
	and $94.6$\% ($95.0$\%) at 300\,K (100\,K), respectively, with an uncertainty of roughly 1\%.
	}
	\begin{tabular}{lcccccc} \hline
		$T$(K) &  $U$(Ba) & $U$(Ce)& $U$(Ir)& $x$(O)    & $U_{11}$(O)& $U_{22}$(O) \\ \hline
		300 &  1299(8) & 740(8) & 475(6) & 0.2592(3) & 1230(140) & 2390(110) \\
		100 & 1087(13) & 738(18)& 585(12)& 0.2590(7) & 1100(300) & 2200(200) 
	\end{tabular}
\end{table}

However, the atomic displacement parameters shown in table I provide evidence for local distortions since they are 
(i) larger than expected for a purely dynamical displacement, 
(ii) very similar at 300\,K and 100\,K, 
and (iii) similar for the heavy Ba ions and the lighter O ions. 
The large values observed for O perpendicular to its bond at room temperature reflect 
the general instability of a perovskite against tilting. But the small difference in the room-temperature 
and 100\,K displacement values in general indicates some local distortions. Moreover, a normal dynamical 
effect cannot explain the fact that the atomic displacement parameter of the heavy Ba is of the same magnitude 
(a root mean square displacement of the order of 0.1\,\AA) as the one of the much lighter O.\@ 
Fits of the data in space groups with the same translation lattice but
reduced symmetry do not yield significant improvement. 
However, a split model in which the Ba ions are statistically distributed over sites slightly 
displaced by $\delta_{\rm Ba}$ against the cubic (0.25,0.25,0.25) 
position results in $\delta_{\rm Ba}$\,=\,0.13(1)\,\AA\ and 0.14(2)\,\AA\ 
at 300\,K and 100\,K, respectively.  
The statistical character may be related to the existence of about 5\,\% of vacancies on the Ir sites. 

One example for structural distortions that were first sensed by enlarged atomic 
	displacement parameters is KH$_2$PO$_4$, a prototype ferroelectric material that exhibits 
	highly enlarged atomic displacement parameters above its ferroelectric transition of 
	order-disorder character \cite{Nelmes1987}. Another example is La$_{1.85}$Sr$_{0.15}$CuO$_4$, 
	in which enhanced atomic displacement factors are observed in samples which do not show 
	long-range tilt order \cite{Braden2001}.
	In \BaCeIrO, the presence of local distortions from cubic symmetry is supported by our RIXS data, 
	see below. This can be reconciled with the observation of a global cubic structure in x-ray 
	diffraction by assuming a negligible correlation length of the distortions.
We conclude that Ba$_2$CeIrO$_6$ is cubic on average but exhibits small local distortions.

This result is in contrast to an earlier report on a monoclinic structure of \BaCeIrO\ 
in Ref.\ \cite{Wakeshima2000} which was based on powder data and a tiny monoclinic distortion 
of the metric ($a/b$\,=\,$1.0016$ \cite{Wakeshima2000}) that can result from the broadening of Bragg peaks. 
Combining the typical rotation of octahedra in the GdFeO$_3$ structure type of a perovskite $AB$O$_3$ 
with the doubling of the unit cell in the double perovskite results in a monoclinic distortion, $P2_1/c$. 
One may examine the possible instability of Ba$_2$CeIrO$_6$ by calculating the Goldschmidt tolerance 
factor for perovskites
\begin{equation}
t_p = \frac{r_A + r_O}{\sqrt{2}(r_B+r_O)}
\label{eq:tolerance}
\end{equation}
with the ionic radii $r_i$. For an ideal cubic perovskite, $t_p$\,=\,1. 
For the hypothetic perovskites BaIrO$_3$ and  BaCeO$_3$ this yields $t_p$\,=\,1.06 and 0.94 not 
indicating sizable bond-length mismatch. The same analysis for distorted Sr$_2$CeIrO$_6$ yields 
$t_p$\,=\,0.90 and 0.80. Alternatively, one may consider the tolerance factor $t_{dp}$ 
for an ordered double perovskite $A_2BB^\prime$O$_6$
with $r_B$ in Eq.\ \eqref{eq:tolerance} to be replaced by $(r_B+r_{B^\prime})/2$.
A mono\-clinic structure is favored for $t_{dp} \leq 0.96$ while values close to 1 point towards a 
cubic structure \cite{Vasala2015}. 
For \BaCeIrO, one finds $t_{dp}$\,=\,$0.991$, supporting a cubic structure.

\section{Magnetic susceptibility}

\begin{figure}[t]
	\centering
	\includegraphics[width=\columnwidth]{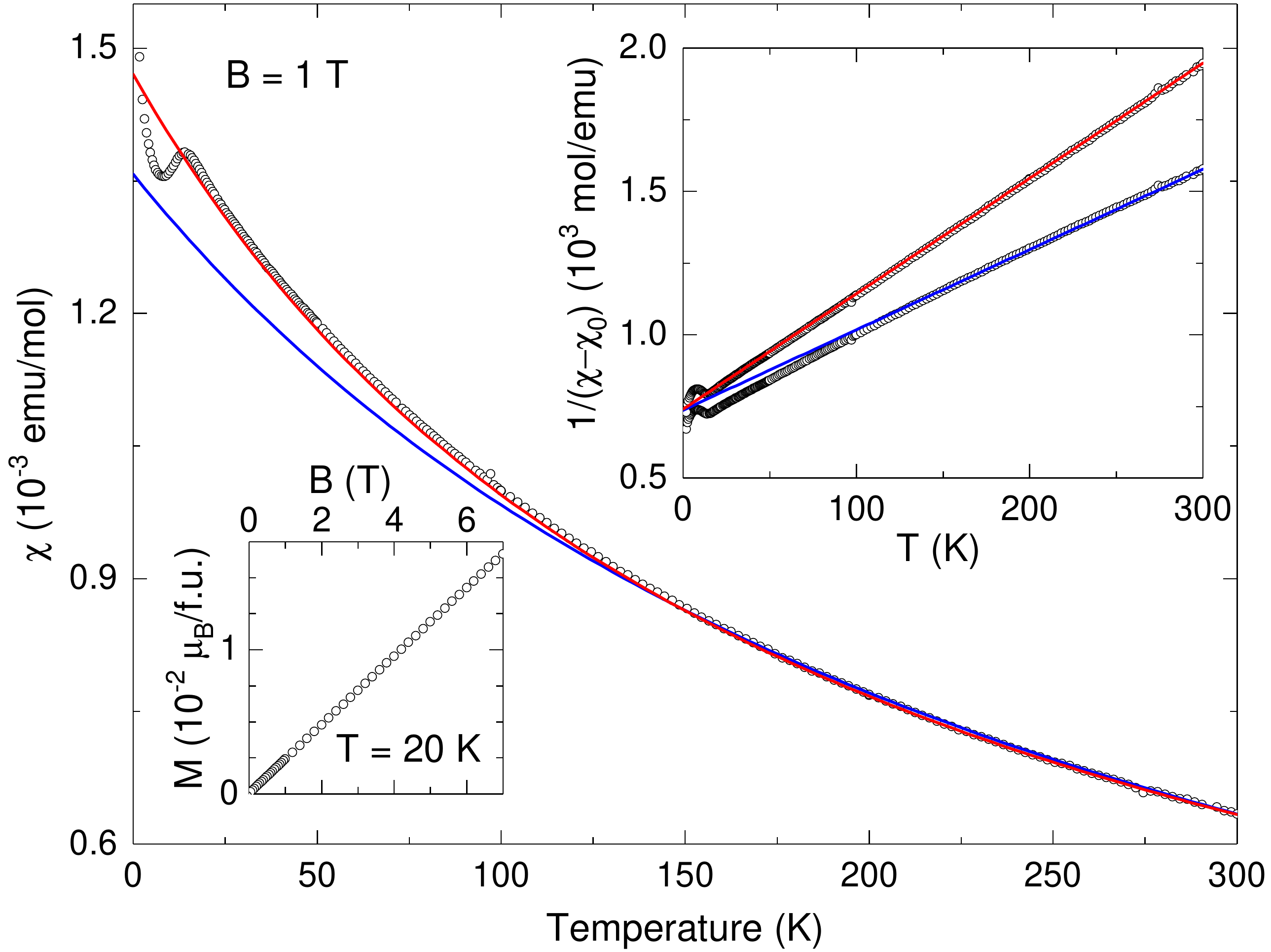}
	\caption{{\bf Magnetic susceptibility of \BaCeIrO,}
	measured on an assembly of 20 single crystals with 21\,mg total mass in a field of 1\,T. 
	The red curve denotes a Curie-Weiss fit which yields $\chi_0$\,=\,$1.2\cdot 10^{-4}$\,emu/mol, 
	while the blue curve corresponds to a fit restricted to $\chi_0$\,=\,$0$.  
	The upper inset shows the same data plotted as $1/(\chi-\chi_0)$. 
	The lower inset displays the linear field dependence of the magnetization, 
	using T\,=\,20\,K as a representative example.
	\label{fig:Susceptibility}
	}
\end{figure}

To explore the magnetism of the local moments in \BaCeIrO\ we measured the magnetization and 
the magnetic susceptibility $\chi(T)$. 
We used an assembly of 20 small single crystals in order to enhance the 
	magnetic signal. The crystals were not aligned because an isotropic magnetic susceptibility 
	is expected in the paramagnetic phase of the (global) cubic structure. 
	As shown in the lower inset of Fig.\ \ref{fig:Susceptibility}, we observe a field-linear 
	magnetization. The main panel of Fig.\ \ref{fig:Susceptibility} shows $\chi(T)$. 
	Its high-temperature behavior 
essentially follows a Curie-Weiss behavior from 300\,K 
down to about $T_N$\,=\,14\,K, where a distinct drop in $\chi(T)$ signals antiferromagnetic ordering. 
For the quantitative analysis we use 
\begin{equation}
\label{eg:CurieW}
\chi(T)=N_A\frac{\mu_{\rm eff}^2}{3k_B(T-\Theta_{\rm CW})}+\chi_0 \, ,
\end{equation}
where $N_A$ and $k_B$ denote Avogadro's and Boltzmann's constant, respectively, $\Theta_{\rm CW}$ 
is the Curie-Weiss temperature, and the constant $\chi_0$\,=\,$\chi_{dia}+\chi_{vV}$ represents core diamagnetism 
$\chi_{dia}\! \simeq \! -1.7\cdot 10^{-4}$\,emu/mol and van Vleck paramagnetism $\chi_{vV}\!>\!0$, 
which are expected to be of the same order of magnitude 
\footnote{
$\chi_{dia}\simeq -1.7\cdot 10^{-4}$\,emu/mol  results from the tabulated values \cite{Bain2008} 
for the ions in Ba$_2$CeIrO$_6$.}. 
A fit based on Eq.\ (\ref{eg:CurieW}) describes the data above $T_N$ very well, 
	see red line in Fig.\ \ref{fig:Susceptibility}, and yields the parameters 
	$\chi_{0}$\,=\,$1.2\cdot 10^{-4}$\,emu/mol, $\mu_{\rm eff}$\,=\,$1.41$\,$\mu_B$, and 
	$\Theta_{\rm CW}$\,=\,$-184$\,K.\@  
	Very similar values, $\mu_{\rm eff}$\,=\,$1.3$\,$\mu_B$ and $\Theta_{\rm CW}$\,=\,$-177$\,K,  
    were reported in a previous study on polycrystalline \BaCeIrO{} \cite{Wakeshima2000}. 
To estimate the reliability of our result,
we compare with a fit assuming $\chi_{0}$\,=\,0, which shows Curie-Weiss behavior above about 120\,K 
(blue line in Fig.\ \ref{fig:Susceptibility}), 
$\mu_{\rm eff}$\,=\,$1.69$\,$\mu_B$ and $\Theta_{\rm CW}$\,=\,$-263$\,K, 
i.e., an even larger value of $|\Theta_{\rm CW}|$.
Thus, both fits result in an effective magnetic moment that is moderately reduced from 
$\mu_{\rm eff}$\,=\,$1.73\,\mu_B$ expected for $j$\,=\,$1/2$ moments in an ideal cubic crystal 
field \cite{Moretti_PRL2014} and indicate substantial frustration with a frustration parameter 
$f$\,=\,$\left|\Theta_{\rm CW}\right|/T_N \! > \! 13$.

\section{RIXS}

\BaCeIrO\ indeed realizes nearly ideal local $j$\,=\,$1/2$ moments, which can be inferred from our RIXS results.  
In cubic symmetry, a single $5d^5$ Ir$^{4+}$ site with a $t_{2g}^5$ configuration is expected to show a local 
$j$\,=\,1/2 ground state and a $j$\,=\,3/2 excited state, the so-called spin-orbit exciton, at 1.5\,$\lambda$ 
with $\lambda$\,=\,0.4-0.5\,eV.\@ 
The effect of a non-cubic crystal field is described by the single-site Hamiltonian 
\begin{equation}
H_{\rm single} = \lambda\, \vec{S}\cdot \vec{L} + \Delta_{\rm CF} L_z^2 \, , 
\label{eq:single}
\end{equation}
which shows a crystal-field splitting of the $j$\,=\,$3/2$ quartet and a mixing of 
$j$\,=\,$1/2$ and $3/2$ wavefunctions in the ground state, 
$
|0\rangle = \alpha \left|\frac{1}{2},\frac{1}{2}\right\rangle + \beta \left|\frac{3}{2},\frac{1}{2}\right\rangle 
$ 
in the $|j,j_z\rangle$ basis. 
With $\alpha$\,=\,$(\sin\theta+\sqrt{2}\cos\theta)/\sqrt{3}$ and 
$\tan 2\theta$\,=\,$\sqrt{8}/(1-2\Delta_{\rm CF}/\lambda)$ \cite{Jackeli2009}
we can readily infer the ground state wavefunction by measuring $\Delta_{\rm CF}$. 

To do so, we performed RIXS measurements at the Ir $L_3$ edge, the most sensitive probe for 
the corresponding intra-$t_{2g}$ excitations. 
For $\Delta_{\rm CF}/\lambda \ll 1$, the experimentally observed peak splitting amounts to 
$\Delta_{\rm exp}$\,=\,$\frac{2}{3}\Delta_{\rm CF}$. 
Thus far, all experimental results on the spin-orbit exciton in iridates show a finite non-cubic 
crystal-field splitting \cite{Rau2016,Trebst2017,Gretarsson13,KimRIXS14,Rossi17,Liu12,Moretti14}. 
The smallest values $\Delta_{\rm exp}$\,=\,0.11-0.14\,eV were reported for Rb$_2$IrF$_6$, 
Na$_2$IrO$_3$, and Sr$_2$IrO$_4$ \cite{Rossi17,Gretarsson13,KimRIXS14}. 
In Rb$_2$IrF$_6$, F-Ir-F bond angles vary from $87^\circ$ to $93^\circ$ \cite{Rossi17}, 
while Sr$_2$IrO$_4$ shows distorted IrO$_6$ octahedra with Ir-O bond lengths of 1.98-2.06\,\AA\ 
and Ir-O-Ir bond angles of $157^\circ$ \cite{Crawford94}.
Despite the substantial distortions, 
these compounds are widely accepted as realizations of the $j$\,=\,1/2 scenario. 
In contrast, strong deviations from the $j$\,=\,1/2 model are reported for Sr$_3$CuIrO$_6$ and CaIrO$_3$ 
with $\Delta_{\rm exp}$\,=\,0.23\,eV and 0.6\,eV, respectively \cite{Liu12,Moretti14}.

\begin{figure}[b]
	\centering
	\includegraphics[width=\columnwidth]{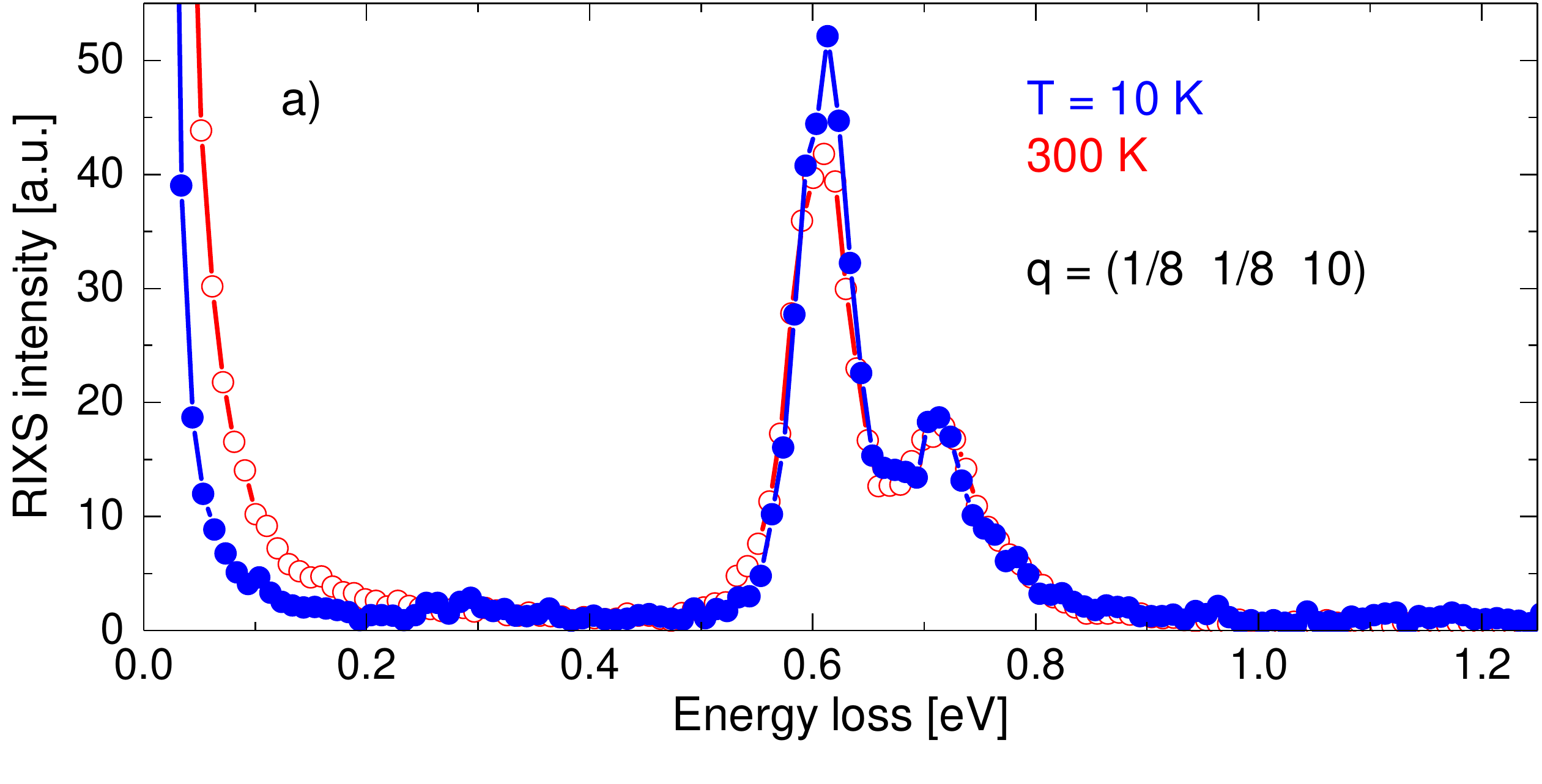}
	\includegraphics[width=\columnwidth]{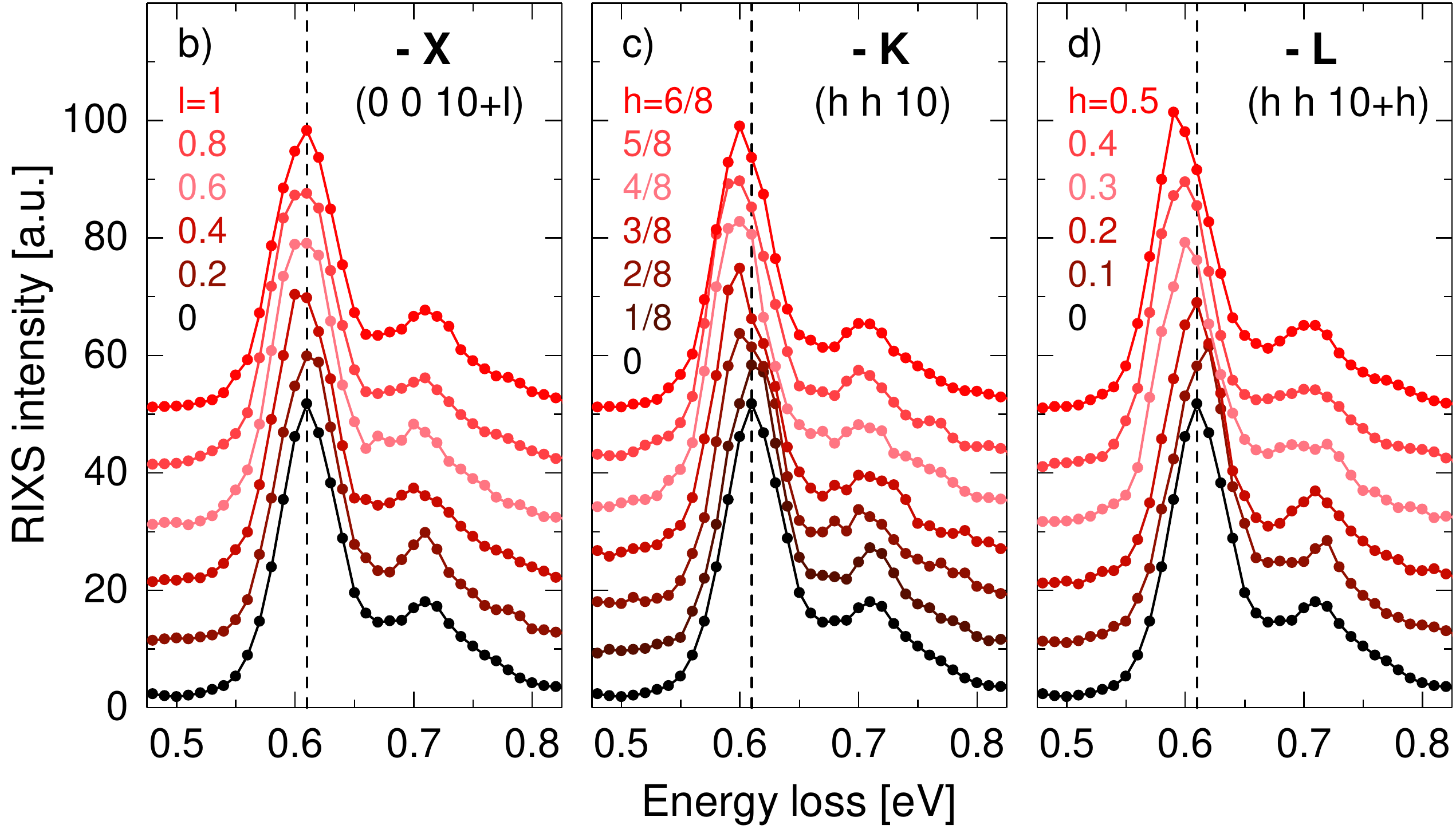}
\caption{{\bf RIXS data of \BaCeIrO.} 
	a) RIXS peaks at 0.61\,eV and 0.71\,eV correspond to excitations to $j$\,=\,$3/2$ states 
	which are split by a non-cubic crystal field. 
	Data at 300\,K show a slightly enhanced peak width but the same splitting as at 10\,K.\@ 
	b)-d) Dispersion along high-symmetry directions at 10\,K.\@ The dashed line	denotes the peak energy 
	at the $\Gamma$ point, 0.61\,eV.\@  
	The largest dispersion is observed from $\Gamma$ to $L$, i.e.\ along ($h$ $h$ $h$). 
	All RIXS spectra were measured in the Brillouin zone around (0 0 10) to achieve a scattering angle $2\theta$ 
	close to $90^\circ$ which suppresses the contribution of the elastic line at zero energy loss. 
	\label{fig:RIXS}
	}
\end{figure}

For \BaCeIrO, we measured RIXS data on a polished \mbox{(0 0 1)} surface at the ID20 beamline 
at ESRF using an incident energy of 11.215\,keV with an overall resolution of 25\,meV \cite{Moretti13,Moretti18}. 
The incident photons were $\pi$ polarized. 
Our data offer a textbook example of the spin-orbit exciton 
by showing two narrow RIXS peaks on a negligible background, see Fig.~\ref{fig:RIXS}. 
Similar RIXS spectra with a slightly larger peak splitting were reported for Rb$_2$IrF$_6$ \cite{Rossi17} 
and Ba$_3$Ti$_{2.7}$Ir$_{0.3}$O$_9$ \cite{Revelli2019}, two compounds with well separated Ir$^{4+}$ ions. 
In comparison, $5d^5$ iridates with stronger hopping such as \NaIrO\ and \SrIrO\ show more complex 
RIXS features \cite{Gretarsson13,KimRIXS14} with, e.g., further peaks, broader line widths, 
and/or a continuum contribution. 
In \BaCeIrO, the peaks are located
at about 0.61\,eV and 0.71\,eV, both at 10\,K and at 300\,K.\@ The observation of {\em two} peaks 
signals non-cubic local distortions in agreement with our analysis of the x-ray diffraction data. 
A fit using two peaks with the Pearson VII line shape \cite{Wang05Pearson} 
that mimics a convolution of an intrinsic Lorentzian line shape and a Gaussian profile with 
the experimental resolution yields a splitting $\Delta_{\rm exp}$\,=\,$(100\pm 4)$\,meV, 
the smallest splitting reported thus far in $L$ edge RIXS for the spin-orbit exciton in iridates 
\cite{Rossi17,Gretarsson13,KimRIXS14,Liu12,Moretti14,Revelli2019}. 
The peak values of 0.61\,eV and 0.71\,eV allow for two different solutions of Eq.\ (\ref{eq:single})
with $\lambda$\,=\,0.43\,eV and $\Delta_{\rm CF}$\,=\,0.17\,eV or -0.15\,eV, which correspond to 
elongation or compression, respectively. This results in a ground-state wavefunction 
\begin{equation}
|0\rangle = 0.991 \left|\frac{1}{2},\frac{1}{2}\right\rangle - 0.130 \left|\frac{3}{2},\frac{1}{2}\right\rangle 
\label{eq:gs}
\end{equation}
in the $|j,j_z\rangle$ basis for elongation, while for compression the coefficients are 0.995 and 0.100, 
respectively. Note that both solutions deviate by less than 1\,\% from the ideal $j$\,=\,1/2 case.

To probe the intersite hopping interactions, we have measured the dispersion via RIXS for 
${\bf q}$ along 
different high-symmetry directions. 
Data along $\Gamma$-$K$ and $\Gamma$-$L$ paths reveal a finite dispersion of up to 15-20\,meV, 
while peak energies are nearly independent of ${\bf q}$ along $\Gamma$-$X$, 
see the lower panels of Fig.~\ref{fig:RIXS}.  
The corresponding delocalization of the $j$\,=\,$3/2$ excited state is a clear signature of 
microscopic hopping processes and intersite interactions that are closely related to the 
magnetic exchange interactions between localized $j$\,=\,1/2 moments \cite{KimRIXS12,KimRIXS14}. 
Roughly, this common microscopic origin is reflected in the common energy scale 
	of 15-20\,meV of the spin-orbit exciton dispersion and the Curie-Weiss temperature, 
	which also is a measure of the size of magnetic exchange interactions.

\begin{figure*}[th]
	\centering
	\includegraphics[width=\linewidth]{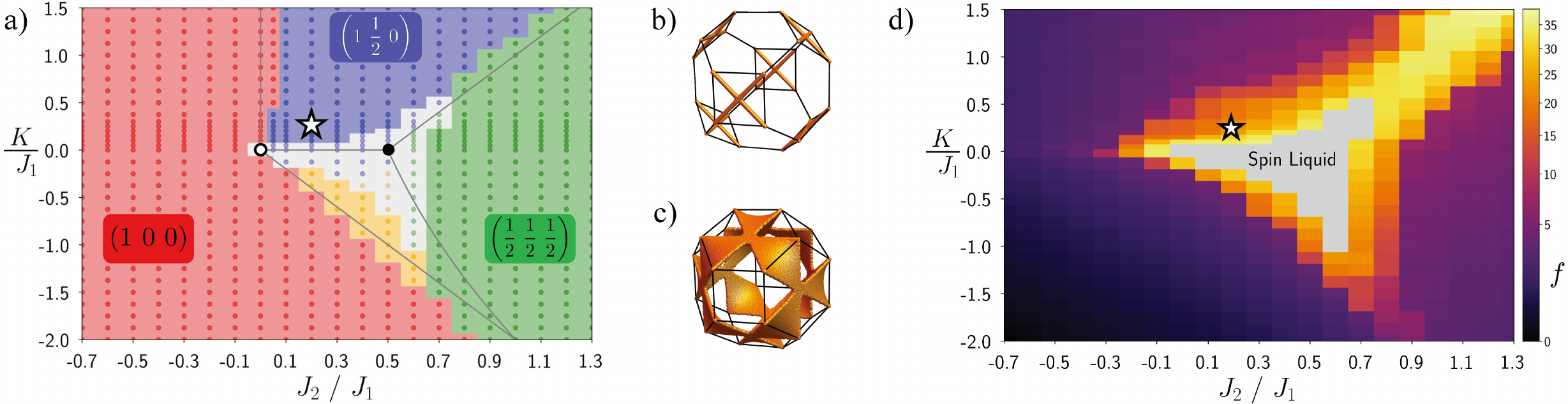}
	\caption{(a) {\bf Phase diagram} for the $J_1$-$J_2$-$K$ model. 
	We find ordered phases with ordering wavevectors $(100)$ (red), $(\frac{1}{2} \frac{1}{2} \frac{1}{2})$ 
	(green), and $(1 \frac{1}{2} 0)$ (blue), an incommensurate spiral phase (yellow) 
	whose ordering wavevector continuously varies within the phase, and a spin liquid regime (gray). 
	The lines indicate the phase boundaries of the classical model for comparison.
	White and black circles mark high-degeneracy points of the classical model (see text), 
    their corresponding sets of ${\bf q}$ vectors are shown in b) and c), respectively. 
	The star indicates the parameter set obtained for \BaCeIrO.
	(d) The frustration parameter $f$\,=\,$|\Theta_{\rm CW}|/T_N$ shows the suppression of ordering tendencies 
	caused by the interplay of geometric and exchange frustration. 
	Gray: spin liquid regime.	
\label{fig:J1J2KModel}
	}
\end{figure*}

\section{Microscopic Model}

\subsection{{\em fcc} lattice with cubic site symmetry}

A symmetry analysis \cite{Cook2015,Aczel2016,Gang2017}  of exchange interactions on the undistorted 
{\em fcc} lattice shows that the most general nearest-neighbor spin Hamiltonian allows for 
Heisenberg coupling $J_1$, Kitaev coupling $K$, and symmetric off-diagonal exchange $\Gamma$. 
We estimate the coupling constants using density functional theory (GGA+U+SOC) for different 
magnetic configurations and $t/U$ perturbation theory for an effective tight-binding model (see Appendix B).
Both approaches consistently yield an antiferromagnetic $J_1 \! \approx \! 5-7$\,meV 
and two subdominant couplings $K \! \approx J_2 \! \approx\! 0.2\,J_1$, where $J_2$ denotes a 
next-nearest neighbor Heisenberg coupling. We find that $\Gamma/J_1 \! \lesssim \! 0.05$ is negligible. 
The corresponding Curie-Weiss temperature 
$\Theta_{\rm CW}$\,=\,$- (3 J_1 + K + 3 J_2/2) \! \approx \! -200$\,K to $-280$\,K agrees 
with the experimental $\chi(T)$; see Fig.~\ref{fig:Susceptibility}. 
Note that we find an antiferromagnetic Kitaev coupling, in contrast to the ferromagnetic ones 
inferred for the honeycomb-based iridates and $\alpha$-RuCl$_3$ \cite{Winter2016}.  
The ferromagnetic Kitaev coupling of the latter arises from Hund's coupling in the 
virtually excited intermediate state with two holes on the same site, favoring parallel hole spins. 
For the honeycomb materials with a $90^\circ$ Ir-O-Ir exchange path, this translates into a ferromagnetic 
coupling of $j$\,=\,1/2 pseudo-spins. 
In \BaCeIrO, exchange proceeds via an Ir-O-O-Ir path with a different combination of orbitals in the 
virtual state. Again, Hund's coupling favors parallel spins of the two holes, 
but for the relevant orbitals this translates to \textit{antiferromagnetic} 
coupling of $j$\,=\,1/2 pseudo-spins 
(see Appendix C).

To study the competition of geometric and exchange frustration, we explore the minimal microscopic model
\begin{equation}
	\mathcal{H} = J_1 \sum_{\langle i,j \rangle} {\vec {\cal S}}_i \cdot {\vec {\cal S}}_j 
	+ K \sum_{\langle i,j \rangle_\gamma} {\cal S}^\gamma_i  {\cal S}^\gamma_j
				+ J_2 \sum_{\langle\langle i,j \rangle\rangle} {\vec {\cal S}}_i \cdot {\vec {\cal S}}_j ,
	\label{eq:model}
\end{equation}
where $\la i,j\rangle_\gamma$ denotes nearest-neighbor pairs in the plane perpendicular to 
axis $\gamma$\,(=\,$x$,$y$,$z$), 
$\la\la i,j\rangle\rangle$ runs over next-nearest-neighbor pairs, 
and the spin operators ${\vec {\cal S}}$ refer to $j$\,=\,$1/2$ moments. 
We have calculated its rich phase diagram using a pseudofermion functional renormalization group (pf-FRG) 
approach \cite{Reuther2010}. This numerical scheme combines elements from $1/S$ expansion \cite{Baez2016} 
and $1/N$ expansion \cite{Buessen2018,Roscher2018}, allowing it to capture both magnetic order 
and spin-liquid ground states. 
There are four magnetically ordered phases, one of them showing incommensurate spiral order, 
see Fig.~\ref{fig:J1J2KModel}a). 
These phases can be readily understood in the classical limit of model \eqref{eq:model} via 
a Luttinger-Tisza approach \cite{Luttinger1946,Luttinger1951}, with the classical phase boundaries also 
indicated in Fig.~\ref{fig:J1J2KModel}a). 
The quantum model additionally exhibits a spin-liquid phase with no magnetic order. 
Its origin is revealed by two points of special interest in the classical model, 
see white and black circles in Fig.~\ref{fig:J1J2KModel}a): 
(i) $J_2$\,=\,$K$\,=\,$0$, the {\em fcc} nearest-neighbor Heisenberg antiferromagnet. 
It exhibits a degenerate manifold of coplanar spin spiral ground states \cite{Henley1987}.  
The corresponding set of ${\bf q}$ vectors is shown in Fig.~\ref{fig:J1J2KModel}b).
(ii) $J_2$\,=\,$J_1/2$, $K$\,=\,$0$, where three ordered phases meet in the classical model. 
This point features an even larger set of degenerate coplanar spin-spiral ground states, 
depicted by the {\em surface} of ${\bf q}$ vectors in Fig.~\ref{fig:J1J2KModel}c). 
The presence of a considerable (but still subextensive) manifold of (nearly) degenerate low-energy states appears 
to give rise to an extended spin liquid regime in the quantum model, centered around the 
classical high-degeneracy point 
\footnote{
A similar scenario has recently been discussed in the context of the $J_1-J_2$ Heisenberg model 
on the diamond lattice \cite{Bergman2007,Buessen2018b}.
}.

To further investigate the interplay of geometric and exchange frustration, we calculate \cite{Reuther2011c} 
the dimensionless frustration parameter $f$\,=\,$|\Theta_{\rm CW}| / T_N$, see Fig.~\ref{fig:J1J2KModel}d), 
using estimates of $\Theta_{\rm CW}$ and $T_N$ obtained from fits of the magnetic susceptibility numerically 
obtained by FRG calculations.
The frustration parameter diverges in the spin liquid regime due to the absence of finite-temperature order. 
Furthermore, $f$ is particularly large along the phase boundary between the $(1 \frac{1}{2} 0)$ and 
$(\frac{1}{2} \frac{1}{2} \frac{1}{2})$ phases, where both $J_2$ and $K$ are substantial and antiferromagnetic. 
This boosts $|\Theta_{\rm CW}|$ while $T_N$ is small close to the phase boundary. 
Close to the spin-liquid regime for the parameter set estimated for \BaCeIrO\ (cf.\ star in 
Fig.~\ref{fig:J1J2KModel}d)), we also find large values of $f$. 
However, moving away from the spin-liquid regime the frustration is quickly reduced with increasing strength 
of the Kitaev coupling. 
This is consistent with a previous classical Monte Carlo study \cite{Cook2015,Aczel2016},
although such a classical analysis by itself is not reliable in the deep quantum limit of $j\!=\!1/2$. 
Our results show that the Kitaev coupling, in competition with the geometric frustration of the 
Heisenberg exchange, indeed induces 
magnetic order for the system at hand -- in striking contrast 
to a number of $j$\,=\,$1/2$ materials where the Kitaev coupling is primarily considered 
a source of frustration \cite{Rau2016,Trebst2017}.

\subsection{Distortions}

The strong frustration in \BaCeIrO\ boosts the importance of magneto-elastic coupling. 
We find theoretically that even small local distortions severely affect the exchange couplings,
although the ground state wavefunction remains close to the $j$\,=\,$1/2$ limit, see Eq.\ \eqref{eq:gs}. 
The precise character of the local distortions cannot be determined from our x-ray diffraction 
results, which show global cubic symmetry.
A putative tetragonal distortion of strength $\Delta_{\rm CF}$ gives rise to a strong spatial anisotropy,
which can be rationalized as follows.
Focusing, e.g., on the dominant contribution to exchange within the $xy$ plane, we find 
$J_1^{xy}$ to depend quadratically on the occupation probability of the $xy$ orbital.
Comparing cubic $\Delta_{\rm CF}$\,=\,$0$ with the distorted case $\Delta_{\rm CF}/\lambda \approx 0.4$
derived above, the $xy$ occupation is strongly enhanced from $1/3$ to $0.46$ 
and as a result the nearest-neighbor Heisenberg exchange $J_1^{xy}$ increases by about a factor of 
two, which corresponds to a dramatic magneto-elastic effect.  
In particular, $\Delta_{\rm CF}\! >\! 0$ strengthens (weakens) $J_1$, $J_2$, and $K$ 
in the $xy$ plane ($yz$ and $xz$ planes), while $\Delta_{\rm CF}\! < \! 0$ has the reverse effect. 
This strong spatial anisotropy of the couplings is sketched in Fig.~\ref{fig:TetragonalDistortion}c). 
Note that a change of the $xy$ occupation $\sin^2 \theta$ has a much more pronounced effect 
on the exchange, $J_1 \propto \sin^4 \theta$, than on the coefficient 
$\alpha$\,=\,$\sqrt{1/3}\sin\theta+\sqrt{2/3}\sqrt{1-\sin^2\theta}$ 
of the $|\frac{1}{2},\frac{1}{2}\rangle$ contribution to the ground state wavefunction. 
The comparably small change of $\alpha$ indicates a small deviation from a cubic charge distribution and a 
concomitant small energy cost for lattice distortions, while the larger change of $J_1$ and in particular 
its spatial anisotropy yield a significant gain of magnetic energy, particularly in the presence of frustration. 

\begin{figure}[t]
	\centering
	\includegraphics[width=\columnwidth]{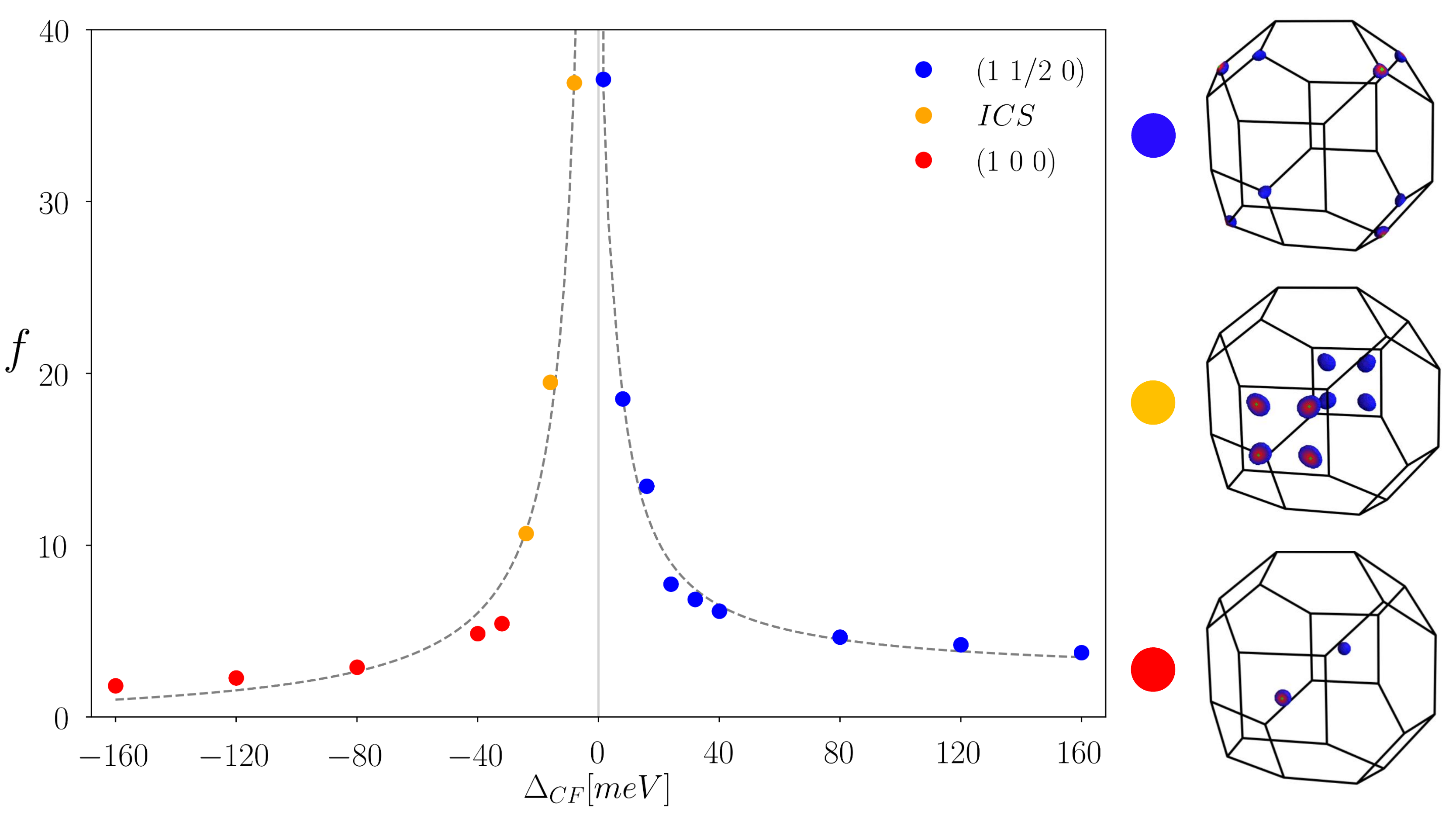}
	\caption{Frustration parameter as a function of the tetragonal distortion $\Delta_{\rm CF}$. 
		Colors indicate different types of magnetic order, with respective structure factors, 
		calculated using the pf-FRG scheme, shown to the right. 
		The abbreviation ICS denotes the incommensurate spiral phase.
}
	\label{fig:TetragonalDistortionFrustrationParameters}
\end{figure}

To analyze the effect of a {\em global} tetragonal lattice distortion, we have simulated 
a variant of the $J_1$-$J_2$-$K$ model, cf.\ Eq.~(5), with anisotropic coupling strengths, enhancing/reducing 
the couplings as described above and illustrated in Fig.~1c). 
Specifically, we have modeled the dependence of the coupling parameters on the distortion $\Delta_{\rm CF}$ as 
\begin{equation}
J_1 \to  \left\{ \begin{array}{l}
J_1 + \delta J (\Delta_{\rm CF}), \hspace{0.84cm}  xy \,\, {\rm plane},\\
J_1 - \delta J (\Delta_{\rm CF}), \hspace{0.25cm} yz,\, xz \,\, {\rm planes},
\end{array} \right.
\end{equation}
and, analogously, for $J_2$ and $K$.
For $\Delta_{\rm CF}$\,=\,$0$ this corresponds to the parameter set indicated by a star in Fig.\ 4d),
i.e. $J_2$\,=\,$K$\,=\,$0.2 J_1$, while for $\Delta_{\rm CF}$\,=\,$\Delta^{\rm max}_{\rm CF}$\,=\,$160$\,meV 
we have the enhanced $xy$ couplings 
$J_{1}^{xy}$\,=\,$2 J_{1}$, $J_2^{xy}$\,=\,$1.2 J_{2}$, and $K^{xy}$\,=\,$1.2 K$ 
for simultaneously reduced parameters
$J_{1}^{xz, yz}$\,=\,$0.6 J_{1}$, $J_2^{xz, yz}$\,=\,$0.7 J_{2}$, and $K^{xz, yz}$\,=\,$0.8 K$
in the $xz$ and $yz$ plane.
The  function $\delta J (\Delta_{\rm CF})$ is well-approximated as a linear interpolation
\begin{equation}
\delta J (\Delta_{\rm CF}) = \frac{\Delta_{\rm CF}}{\Delta^{\rm max}_{\rm CF}} \big{(} J(\Delta^{\rm max}_{\rm CF}) - J(\Delta_{\rm CF}=0) \big{)} \,.
\end{equation}
Results for pf-FRG calculations are summarized in Fig.~\ref{fig:TetragonalDistortionFrustrationParameters} 
showing the frustration parameter $f = |\Theta_{\rm CW}|/T_{\rm N}$ as a function of the distortion 
$\Delta_{\rm CF}$.
As clearly visible, the frustration is strongly suppressed by the distortions, it quickly approaches 
a non-frustrated regime $f \lesssim 5$ for distortions of the order of $| \Delta_{\rm CF} | \approx 40$\,meV, 
independent of the sign of the distortion.
Additionally, the data indicate a potential change of magnetic order, for instance to (100) order, 
depending on the sign and the strength of $\Delta_{\rm CF}$. These results agree 
with the small $f \lesssim 2$ reported for the globally distorted monoclinic $5d^5$ 
double perovskites La$_2$ZnIrO$_6$ and La$_2$MgIrO$_6$ \cite{Aczel2016}. 
However, this result for a {\em global} distortion is significantly smaller than the value of 
$f \! > \! 13$ measured in \BaCeIrO. 
This suggests that the statistical distribution of {\em local} distortions along different tetragonal axes, 
in contrast to a global distortion, is important in order to recover the experimentally observed large frustration.

The existence of a weak (but unavoidable) magneto-elastic effect was recently discussed for 
Sr$_2$IrO$_4$ \cite{Khaliullin2018}. The strong effect of the magneto-elastic coupling 
in \BaCeIrO, however, is due to an additional mechanism arising from an interplay of distortions 
and magnetic frustration, which is not present in Sr$_2$IrO$_4$, but will be relevant, e.g., 
in tetragonal bilayer Sr$_3$Ir$_2$O$_7$ and in the 3D honeycomb iridates.

\section{Conclusions}

The spin-orbit entangled $j$\,=\,1/2  wavefunction has proved to be a versatile source for novel states 
of quantum matter. Its experimental realization in the double perovskite \BaCeIrO\ deviates less than 1\,\%
from the ideal theoretical scenario for a cubic system and is one of the most pristine 
$j$\,=\,1/2 incarnations reported so far in the literature. 
Combining structural analysis, magnetic susceptibility measurements, and RIXS with \textit{ab initio} 
and functional renormalization group calculations for the obtained microscopic model Hamiltonian 
we find that the collective magnetism of this {\em fcc} compound is governed by a competition of 
geometrical frustration, Kitaev-type bond-directional exchange, and magneto-elastic coupling. 
In striking contrast to the honeycomb-based Kitaev materials, the Kitaev exchange is 
antiferromagnetic and in fact stabilizes long-range magnetic order in proximity to a spin liquid phase. 
Importantly, the exchange couplings turn out to be highly sensitive to small deviations from cubic symmetry, 
giving rise to a dramatic magneto-elastic coupling. 
This should be contrasted with the common notion that $j$\,=\,1/2 moments are not Jahn-Teller active,
as the orbital degeneracy is lifted by spin-orbit coupling.
The strong magneto-elastic coupling resurrects the prominent role of lattice distortions on the low-energy properties 
of $j$\,=\,1/2 compounds. 
\\

\textbf{Note added:} 
After submission of our manuscript, Aczel \textit{et al.} reported similar experimental results on polycrystalline 
samples of \BaCeIrO\ \cite{Aczel2019}. They find (100) magnetic order, in agreement with our calculations for 
negative $\Delta_{\rm CF}$. 
Furthermore, Khan \textit{et al.} \cite{Khan2019} reported on the realization of cubic site symmetry in 
K$_2$IrCl$_6$ based on x-ray diffraction data; a spectroscopic proof of this claim is still missing.

\acknowledgments

We acknowledge funding from the Deutsche Forschungsgemeinschaft (DFG, German Research Foundation) 
-- Project numbers 277146847 and 247310070 -- CRC 1238 (projects A02, B01, B02, B03, C02, C03) 
and CRC 1143 (project A05), respectively. 
M.H.\@ acknowledges partial funding by the Knut and Alice Wallenberg Foundation and the Swedish Research Council. 
A.P.\@ acknowledges support from NSERC of Canada and the Canadian Institute for Advanced Research, and the
support and hospitality of the University of Cologne.
The numerical simulations were performed on the JURECA booster at FZ J\"ulich and the CHEOPS cluster at RRZK Cologne. 
DFT calculations were supported by the Russian Science Foundation via project 17-12-01207.


\begin{thebibliography}{63}%
	\makeatletter
	\providecommand \@ifxundefined [1]{%
		\@ifx{#1\undefined}
	}%
	\providecommand \@ifnum [1]{%
		\ifnum #1\expandafter \@firstoftwo
		\else \expandafter \@secondoftwo
		\fi
	}%
	\providecommand \@ifx [1]{%
		\ifx #1\expandafter \@firstoftwo
		\else \expandafter \@secondoftwo
		\fi
	}%
	\providecommand \natexlab [1]{#1}%
	\providecommand \enquote  [1]{``#1''}%
	\providecommand \bibnamefont  [1]{#1}%
	\providecommand \bibfnamefont [1]{#1}%
	\providecommand \citenamefont [1]{#1}%
	\providecommand \href@noop [0]{\@secondoftwo}%
	\providecommand \href [0]{\begingroup \@sanitize@url \@href}%
	\providecommand \@href[1]{\@@startlink{#1}\@@href}%
	\providecommand \@@href[1]{\endgroup#1\@@endlink}%
	\providecommand \@sanitize@url [0]{\catcode `\\12\catcode `\$12\catcode
		`\&12\catcode `\#12\catcode `\^12\catcode `\_12\catcode `\%12\relax}%
	\providecommand \@@startlink[1]{}%
	\providecommand \@@endlink[0]{}%
	\providecommand \url  [0]{\begingroup\@sanitize@url \@url }%
	\providecommand \@url [1]{\endgroup\@href {#1}{\urlprefix }}%
	\providecommand \urlprefix  [0]{URL }%
	\providecommand \Eprint [0]{\href }%
	\providecommand \doibase [0]{http://dx.doi.org/}%
	\providecommand \selectlanguage [0]{\@gobble}%
	\providecommand \bibinfo  [0]{\@secondoftwo}%
	\providecommand \bibfield  [0]{\@secondoftwo}%
	\providecommand \translation [1]{[#1]}%
	\providecommand \BibitemOpen [0]{}%
	\providecommand \bibitemStop [0]{}%
	\providecommand \bibitemNoStop [0]{.\EOS\space}%
	\providecommand \EOS [0]{\spacefactor3000\relax}%
	\providecommand \BibitemShut  [1]{\csname bibitem#1\endcsname}%
	\let\auto@bib@innerbib\@empty
	\bibitem [{\citenamefont {Witczak-Krempa}\ \emph {et~al.}(2014)\citenamefont
		{Witczak-Krempa}, \citenamefont {Chen}, \citenamefont {Kim},\ and\
		\citenamefont {Balents}}]{WitczakKrempa2014}%
	\BibitemOpen
	\bibfield  {author} {\bibinfo {author} {\bibfnamefont {W.}~\bibnamefont
			{Witczak-Krempa}}, \bibinfo {author} {\bibfnamefont {G.}~\bibnamefont
			{Chen}}, \bibinfo {author} {\bibfnamefont {Y.~B.}\ \bibnamefont {Kim}}, \
		and\ \bibinfo {author} {\bibfnamefont {L.}~\bibnamefont {Balents}},\
	}\bibfield  {title} {\enquote {\bibinfo {title} {{Correlated Quantum
					Phenomena in the Strong Spin-Orbit Regime}},}\ }\href {\doibase
		10.1146/annurev-conmatphys-020911-125138} {\bibfield  {journal} {\bibinfo
			{journal} {Annu. Rev. Condens. Matter Phys.}\ }\textbf {\bibinfo {volume}
			{5}},\ \bibinfo {pages} {57} (\bibinfo {year} {2014})}\BibitemShut {NoStop}%
	\bibitem [{\citenamefont {Rau}\ \emph {et~al.}(2016)\citenamefont {Rau},
		\citenamefont {Lee},\ and\ \citenamefont {Kee}}]{Rau2016}%
	\BibitemOpen
	\bibfield  {author} {\bibinfo {author} {\bibfnamefont {J.~G.}\ \bibnamefont
			{Rau}}, \bibinfo {author} {\bibfnamefont {E.~Kin-Ho}\ \bibnamefont {Lee}}, \
		and\ \bibinfo {author} {\bibfnamefont {H.-Y.}\ \bibnamefont {Kee}},\
	}\bibfield  {title} {\enquote {\bibinfo {title} {{Spin-Orbit Physics Giving
					Rise to Novel Phases in Correlated Systems: Iridates and Related
					Materials}},}\ }\href {\doibase 10.1146/annurev-conmatphys-031115-011319}
	{\bibfield  {journal} {\bibinfo  {journal} {Annu. Rev. Condens. Matter
				Phys.}\ }\textbf {\bibinfo {volume} {7}},\ \bibinfo {pages} {195--221}
		(\bibinfo {year} {2016})}\BibitemShut {NoStop}%
	\bibitem [{\citenamefont {Khaliullin}(2005)}]{Khaliullin2005}%
	\BibitemOpen
	\bibfield  {author} {\bibinfo {author} {\bibfnamefont {G.}~\bibnamefont
			{Khaliullin}},\ }\bibfield  {title} {\enquote {\bibinfo {title} {{Orbital
					Order and Fluctuations in Mott Insulators}},}\ }\href {\doibase
		10.1143/PTPS.160.155} {\bibfield  {journal} {\bibinfo  {journal} {Prog.
				Theor. Phys. Suppl.}\ }\textbf {\bibinfo {volume} {160}},\ \bibinfo {pages}
		{155} (\bibinfo {year} {2005})}\BibitemShut {NoStop}%
	\bibitem [{\citenamefont {Chen}\ and\ \citenamefont
		{Balents}(2008)}]{Chen2008}%
	\BibitemOpen
	\bibfield  {author} {\bibinfo {author} {\bibfnamefont {G.}~\bibnamefont
			{Chen}}\ and\ \bibinfo {author} {\bibfnamefont {L.}~\bibnamefont {Balents}},\
	}\bibfield  {title} {\enquote {\bibinfo {title} {{Spin-orbit effects in
					${\text{Na}}_{4}{\text{Ir}}_{3}{\text{O}}_{8}$: A hyper-kagome lattice
					antiferromagnet}},}\ }\href {\doibase 10.1103/PhysRevB.78.094403} {\bibfield
		{journal} {\bibinfo  {journal} {Phys. Rev. B}\ }\textbf {\bibinfo {volume}
			{78}},\ \bibinfo {pages} {094403} (\bibinfo {year} {2008})}\BibitemShut
	{NoStop}%
	\bibitem [{\citenamefont {Jackeli}\ and\ \citenamefont
		{Khaliullin}(2009)}]{Jackeli2009}%
	\BibitemOpen
	\bibfield  {author} {\bibinfo {author} {\bibfnamefont {G.}~\bibnamefont
			{Jackeli}}\ and\ \bibinfo {author} {\bibfnamefont {G.}~\bibnamefont
			{Khaliullin}},\ }\bibfield  {title} {\enquote {\bibinfo {title} {{Mott
					Insulators in the Strong Spin-Orbit Coupling Limit: From Heisenberg to a
					Quantum Compass and Kitaev Models}},}\ }\href {\doibase
		10.1103/PhysRevLett.102.017205} {\bibfield  {journal} {\bibinfo  {journal}
			{Phys. Rev. Lett.}\ }\textbf {\bibinfo {volume} {102}},\ \bibinfo {pages}
		{017205} (\bibinfo {year} {2009})}\BibitemShut {NoStop}%
	\bibitem [{\citenamefont {Kim}\ \emph {et~al.}(2014{\natexlab{a}})\citenamefont
		{Kim}, \citenamefont {Krupin}, \citenamefont {Denlinger}, \citenamefont
		{Bostwick}, \citenamefont {Rotenberg}, \citenamefont {Zhao}, \citenamefont
		{Mitchell}, \citenamefont {Allen},\ and\ \citenamefont {Kim}}]{Kim2014}%
	\BibitemOpen
	\bibfield  {author} {\bibinfo {author} {\bibfnamefont {Y.~K.}\ \bibnamefont
			{Kim}}, \bibinfo {author} {\bibfnamefont {O.}~\bibnamefont {Krupin}},
		\bibinfo {author} {\bibfnamefont {J.~D.}\ \bibnamefont {Denlinger}}, \bibinfo
		{author} {\bibfnamefont {A.}~\bibnamefont {Bostwick}}, \bibinfo {author}
		{\bibfnamefont {E.}~\bibnamefont {Rotenberg}}, \bibinfo {author}
		{\bibfnamefont {Q.}~\bibnamefont {Zhao}}, \bibinfo {author} {\bibfnamefont
			{J.~F.}\ \bibnamefont {Mitchell}}, \bibinfo {author} {\bibfnamefont {J.~W.}\
			\bibnamefont {Allen}}, \ and\ \bibinfo {author} {\bibfnamefont {B.~J.}\
			\bibnamefont {Kim}},\ }\bibfield  {title} {\enquote {\bibinfo {title} {{Fermi
					arcs in a doped pseudospin-1/2 Heisenberg antiferromagnet}},}\ }\href
	{\doibase 10.1126/science.1251151} {\bibfield  {journal} {\bibinfo  {journal}
			{Science}\ }\textbf {\bibinfo {volume} {345}},\ \bibinfo {pages} {187}
		(\bibinfo {year} {2014}{\natexlab{a}})}\BibitemShut {NoStop}%
	\bibitem [{\citenamefont {de~la Torre}\ \emph {et~al.}(2015)\citenamefont
		{de~la Torre}, \citenamefont {McKeown~Walker}, \citenamefont {Bruno},
		\citenamefont {Ricc\'o}, \citenamefont {Wang}, \citenamefont
		{Gutierrez~Lezama}, \citenamefont {Scheerer}, \citenamefont {Giriat},
		\citenamefont {Jaccard}, \citenamefont {Berthod}, \citenamefont {Kim},
		\citenamefont {Hoesch}, \citenamefont {Hunter}, \citenamefont {Perry},
		\citenamefont {Tamai},\ and\ \citenamefont {Baumberger}}]{Torre2015}%
	\BibitemOpen
	\bibfield  {author} {\bibinfo {author} {\bibfnamefont {A.}~\bibnamefont
			{de~la Torre}}, \bibinfo {author} {\bibfnamefont {S.}~\bibnamefont
			{McKeown~Walker}}, \bibinfo {author} {\bibfnamefont {F.~Y.}\ \bibnamefont
			{Bruno}}, \bibinfo {author} {\bibfnamefont {S.}~\bibnamefont {Ricc\'o}},
		\bibinfo {author} {\bibfnamefont {Z.}~\bibnamefont {Wang}}, \bibinfo {author}
		{\bibfnamefont {I.}~\bibnamefont {Gutierrez~Lezama}}, \bibinfo {author}
		{\bibfnamefont {G.}~\bibnamefont {Scheerer}}, \bibinfo {author}
		{\bibfnamefont {G.}~\bibnamefont {Giriat}}, \bibinfo {author} {\bibfnamefont
			{D.}~\bibnamefont {Jaccard}}, \bibinfo {author} {\bibfnamefont
			{C.}~\bibnamefont {Berthod}}, \bibinfo {author} {\bibfnamefont {T.~K.}\
			\bibnamefont {Kim}}, \bibinfo {author} {\bibfnamefont {M.}~\bibnamefont
			{Hoesch}}, \bibinfo {author} {\bibfnamefont {E.~C.}\ \bibnamefont {Hunter}},
		\bibinfo {author} {\bibfnamefont {R.~S.}\ \bibnamefont {Perry}}, \bibinfo
		{author} {\bibfnamefont {A.}~\bibnamefont {Tamai}}, \ and\ \bibinfo {author}
		{\bibfnamefont {F.}~\bibnamefont {Baumberger}},\ }\bibfield  {title}
	{\enquote {\bibinfo {title} {{Collapse of the Mott Gap and Emergence of a
					Nodal Liquid in Lightly Doped ${\mathrm{Sr}}_{2}{\mathrm{IrO}}_{4}$}},}\
	}\href {\doibase 10.1103/PhysRevLett.115.176402} {\bibfield  {journal}
		{\bibinfo  {journal} {Phys. Rev. Lett.}\ }\textbf {\bibinfo {volume} {115}},\
		\bibinfo {pages} {176402} (\bibinfo {year} {2015})}\BibitemShut {NoStop}%
	\bibitem [{\citenamefont {Cao}\ \emph {et~al.}(2016)\citenamefont {Cao},
		\citenamefont {Wang}, \citenamefont {Waugh}, \citenamefont {Reber},
		\citenamefont {Li}, \citenamefont {Zhou}, \citenamefont {Parham},
		\citenamefont {Park}, \citenamefont {Plumb}, \citenamefont {Rotenberg},
		\citenamefont {Bostwick}, \citenamefont {Denlinger}, \citenamefont {Qi},
		\citenamefont {Hermele}, \citenamefont {Cao},\ and\ \citenamefont
		{Dessau}}]{Cao2016a}%
	\BibitemOpen
	\bibfield  {author} {\bibinfo {author} {\bibfnamefont {Y.}~\bibnamefont
			{Cao}}, \bibinfo {author} {\bibfnamefont {Q.}~\bibnamefont {Wang}}, \bibinfo
		{author} {\bibfnamefont {J.~A.}\ \bibnamefont {Waugh}}, \bibinfo {author}
		{\bibfnamefont {T.~J.}\ \bibnamefont {Reber}}, \bibinfo {author}
		{\bibfnamefont {H.}~\bibnamefont {Li}}, \bibinfo {author} {\bibfnamefont
			{X.}~\bibnamefont {Zhou}}, \bibinfo {author} {\bibfnamefont {S.}~\bibnamefont
			{Parham}}, \bibinfo {author} {\bibfnamefont {S.~R.}\ \bibnamefont {Park}},
		\bibinfo {author} {\bibfnamefont {N.~C.}\ \bibnamefont {Plumb}}, \bibinfo
		{author} {\bibfnamefont {E.}~\bibnamefont {Rotenberg}}, \bibinfo {author}
		{\bibfnamefont {A.}~\bibnamefont {Bostwick}}, \bibinfo {author}
		{\bibfnamefont {J.~D.}\ \bibnamefont {Denlinger}}, \bibinfo {author}
		{\bibfnamefont {T.}~\bibnamefont {Qi}}, \bibinfo {author} {\bibfnamefont
			{M.~A.}\ \bibnamefont {Hermele}}, \bibinfo {author} {\bibfnamefont
			{G.}~\bibnamefont {Cao}}, \ and\ \bibinfo {author} {\bibfnamefont {D.~S.}\
			\bibnamefont {Dessau}},\ }\bibfield  {title} {\enquote {\bibinfo {title}
			{{Hallmarks of the Mott-metal crossover in the hole-doped pseudospin-1/2 Mott
					insulator Sr$_2$IrO$_4$}},}\ }\href {http://dx.doi.org/10.1038/ncomms11367}
	{\bibfield  {journal} {\bibinfo  {journal} {Nat. Commun.}\ }\textbf {\bibinfo
			{volume} {7}},\ \bibinfo {pages} {11367} (\bibinfo {year}
		{2016})}\BibitemShut {NoStop}%
	\bibitem [{\citenamefont {Yan}\ \emph {et~al.}(2015)\citenamefont {Yan},
		\citenamefont {Ren}, \citenamefont {Xu}, \citenamefont {Xie}, \citenamefont
		{Tao}, \citenamefont {Choi}, \citenamefont {Lee}, \citenamefont {Choi},
		\citenamefont {Zhang},\ and\ \citenamefont {Feng}}]{Yan2015}%
	\BibitemOpen
	\bibfield  {author} {\bibinfo {author} {\bibfnamefont {Y.~J.}\ \bibnamefont
			{Yan}}, \bibinfo {author} {\bibfnamefont {M.~Q.}\ \bibnamefont {Ren}},
		\bibinfo {author} {\bibfnamefont {H.~C.}\ \bibnamefont {Xu}}, \bibinfo
		{author} {\bibfnamefont {B.~P.}\ \bibnamefont {Xie}}, \bibinfo {author}
		{\bibfnamefont {R.}~\bibnamefont {Tao}}, \bibinfo {author} {\bibfnamefont
			{H.~Y.}\ \bibnamefont {Choi}}, \bibinfo {author} {\bibfnamefont
			{N.}~\bibnamefont {Lee}}, \bibinfo {author} {\bibfnamefont {Y.~J.}\
			\bibnamefont {Choi}}, \bibinfo {author} {\bibfnamefont {T.}~\bibnamefont
			{Zhang}}, \ and\ \bibinfo {author} {\bibfnamefont {D.~L.}\ \bibnamefont
			{Feng}},\ }\bibfield  {title} {\enquote {\bibinfo {title} {{Electron-Doped
					${\mathrm{Sr}}_{2}{\mathrm{IrO}}_{4}$: An Analogue of Hole-Doped Cuprate
					Superconductors Demonstrated by Scanning Tunneling Microscopy}},}\ }\href
	{\doibase 10.1103/PhysRevX.5.041018} {\bibfield  {journal} {\bibinfo
			{journal} {Phys. Rev. X}\ }\textbf {\bibinfo {volume} {5}},\ \bibinfo {pages}
		{041018} (\bibinfo {year} {2015})}\BibitemShut {NoStop}%
	\bibitem [{\citenamefont {Kim}\ \emph {et~al.}(2016)\citenamefont {Kim},
		\citenamefont {Sung}, \citenamefont {Denlinger},\ and\ \citenamefont
		{Kim}}]{Kim2016d}%
	\BibitemOpen
	\bibfield  {author} {\bibinfo {author} {\bibfnamefont {Y.~K.}\ \bibnamefont
			{Kim}}, \bibinfo {author} {\bibfnamefont {N.~H.}\ \bibnamefont {Sung}},
		\bibinfo {author} {\bibfnamefont {J.~D.}\ \bibnamefont {Denlinger}}, \ and\
		\bibinfo {author} {\bibfnamefont {B.~J.}\ \bibnamefont {Kim}},\ }\bibfield
	{title} {\enquote {\bibinfo {title} {{Observation of a d-wave gap in
					electron-doped Sr$_2$IrO$_4$}},}\ }\href
	{http://dx.doi.org/10.1038/nphys3503} {\bibfield  {journal} {\bibinfo
			{journal} {Nat. Phys.}\ }\textbf {\bibinfo {volume} {12}},\ \bibinfo {pages}
		{37} (\bibinfo {year} {2016})}\BibitemShut {NoStop}%
	\bibitem [{\citenamefont {Kim}\ \emph {et~al.}(2008)\citenamefont {Kim},
		\citenamefont {Jin}, \citenamefont {Moon}, \citenamefont {Kim}, \citenamefont
		{Park}, \citenamefont {Leem}, \citenamefont {Yu}, \citenamefont {Noh},
		\citenamefont {Kim}, \citenamefont {Oh}, \citenamefont {Park}, \citenamefont
		{Durairaj}, \citenamefont {Cao},\ and\ \citenamefont {Rotenberg}}]{Kim2008}%
	\BibitemOpen
	\bibfield  {author} {\bibinfo {author} {\bibfnamefont {B.~J.}\ \bibnamefont
			{Kim}}, \bibinfo {author} {\bibfnamefont {H.}~\bibnamefont {Jin}}, \bibinfo
		{author} {\bibfnamefont {S.~J.}\ \bibnamefont {Moon}}, \bibinfo {author}
		{\bibfnamefont {J.-Y.}\ \bibnamefont {Kim}}, \bibinfo {author} {\bibfnamefont
			{B.-G.}\ \bibnamefont {Park}}, \bibinfo {author} {\bibfnamefont {C.~S.}\
			\bibnamefont {Leem}}, \bibinfo {author} {\bibfnamefont {J.}~\bibnamefont
			{Yu}}, \bibinfo {author} {\bibfnamefont {T.~W.}\ \bibnamefont {Noh}},
		\bibinfo {author} {\bibfnamefont {C.}~\bibnamefont {Kim}}, \bibinfo {author}
		{\bibfnamefont {S.-J.}\ \bibnamefont {Oh}}, \bibinfo {author} {\bibfnamefont
			{J.-H.}\ \bibnamefont {Park}}, \bibinfo {author} {\bibfnamefont
			{V.}~\bibnamefont {Durairaj}}, \bibinfo {author} {\bibfnamefont
			{G.}~\bibnamefont {Cao}}, \ and\ \bibinfo {author} {\bibfnamefont
			{E.}~\bibnamefont {Rotenberg}},\ }\bibfield  {title} {\enquote {\bibinfo
			{title} {{Novel ${J}_{\mathrm{eff}}=1/2$ Mott State Induced by Relativistic
					Spin-Orbit Coupling in ${\mathrm{Sr}}_{2}{\mathrm{IrO}}_{4}$}},}\ }\href
	{\doibase 10.1103/PhysRevLett.101.076402} {\bibfield  {journal} {\bibinfo
			{journal} {Phys. Rev. Lett.}\ }\textbf {\bibinfo {volume} {101}},\ \bibinfo
		{pages} {076402} (\bibinfo {year} {2008})}\BibitemShut {NoStop}%
	\bibitem [{\citenamefont {Kim}\ \emph {et~al.}(2009)\citenamefont {Kim},
		\citenamefont {Ohsumi}, \citenamefont {Komesu}, \citenamefont {Sakai},
		\citenamefont {Morita}, \citenamefont {Takagi},\ and\ \citenamefont
		{Arima}}]{Kim2009}%
	\BibitemOpen
	\bibfield  {author} {\bibinfo {author} {\bibfnamefont {B.~J.}\ \bibnamefont
			{Kim}}, \bibinfo {author} {\bibfnamefont {H.}~\bibnamefont {Ohsumi}},
		\bibinfo {author} {\bibfnamefont {T.}~\bibnamefont {Komesu}}, \bibinfo
		{author} {\bibfnamefont {S.}~\bibnamefont {Sakai}}, \bibinfo {author}
		{\bibfnamefont {T.}~\bibnamefont {Morita}}, \bibinfo {author} {\bibfnamefont
			{H.}~\bibnamefont {Takagi}}, \ and\ \bibinfo {author} {\bibfnamefont
			{T.}~\bibnamefont {Arima}},\ }\bibfield  {title} {\enquote {\bibinfo {title}
			{{Phase-Sensitive Observation of a Spin-Orbital Mott State in
					Sr$_2$IrO$_4$}},}\ }\href {\doibase 10.1126/science.1167106} {\bibfield
		{journal} {\bibinfo  {journal} {Science}\ }\textbf {\bibinfo {volume}
			{323}},\ \bibinfo {pages} {1329} (\bibinfo {year} {2009})}\BibitemShut
	{NoStop}%
	\bibitem [{\citenamefont {Trebst}(2017)}]{Trebst2017}%
	\BibitemOpen
	\bibfield  {author} {\bibinfo {author} {\bibfnamefont {S.}~\bibnamefont
			{Trebst}},\ }\bibfield  {title} {\enquote {\bibinfo {title} {{Kitaev
					Materials}},}\ }\href@noop {} {\bibfield  {journal} {\bibinfo  {journal}
			{arXiv:1701.07056}\ } (\bibinfo {year} {2017})}\BibitemShut {NoStop}%
	\bibitem [{\citenamefont {Singh}\ and\ \citenamefont
		{Gegenwart}(2010)}]{Singh2010}%
	\BibitemOpen
	\bibfield  {author} {\bibinfo {author} {\bibfnamefont {Y.}~\bibnamefont
			{Singh}}\ and\ \bibinfo {author} {\bibfnamefont {P.}~\bibnamefont
			{Gegenwart}},\ }\bibfield  {title} {\enquote {\bibinfo {title}
			{{Antiferromagnetic Mott insulating state in single crystals of the honeycomb
					lattice material ${\text{Na}}_{2}{\text{IrO}}_{3}$}},}\ }\href {\doibase
		10.1103/PhysRevB.82.064412} {\bibfield  {journal} {\bibinfo  {journal} {Phys.
				Rev. B}\ }\textbf {\bibinfo {volume} {82}},\ \bibinfo {pages} {064412}
		(\bibinfo {year} {2010})}\BibitemShut {NoStop}%
	\bibitem [{\citenamefont {Singh}\ \emph {et~al.}(2012)\citenamefont {Singh},
		\citenamefont {Manni}, \citenamefont {Reuther}, \citenamefont {Berlijn},
		\citenamefont {Thomale}, \citenamefont {Ku}, \citenamefont {Trebst},\ and\
		\citenamefont {Gegenwart}}]{Singh2012}%
	\BibitemOpen
	\bibfield  {author} {\bibinfo {author} {\bibfnamefont {Y.}~\bibnamefont
			{Singh}}, \bibinfo {author} {\bibfnamefont {S.}~\bibnamefont {Manni}},
		\bibinfo {author} {\bibfnamefont {J.}~\bibnamefont {Reuther}}, \bibinfo
		{author} {\bibfnamefont {T.}~\bibnamefont {Berlijn}}, \bibinfo {author}
		{\bibfnamefont {R.}~\bibnamefont {Thomale}}, \bibinfo {author} {\bibfnamefont
			{W.}~\bibnamefont {Ku}}, \bibinfo {author} {\bibfnamefont {S.}~\bibnamefont
			{Trebst}}, \ and\ \bibinfo {author} {\bibfnamefont {P.}~\bibnamefont
			{Gegenwart}},\ }\bibfield  {title} {\enquote {\bibinfo {title} {{Relevance of
					the Heisenberg-Kitaev Model for the Honeycomb Lattice Iridates
					A$_2$IrO$_3$}},}\ }\href {\doibase 10.1103/PhysRevLett.108.127203} {\bibfield
		{journal} {\bibinfo  {journal} {Phys. Rev. Lett.}\ }\textbf {\bibinfo
			{volume} {108}},\ \bibinfo {pages} {127203} (\bibinfo {year}
		{2012})}\BibitemShut {NoStop}%
	\bibitem [{\citenamefont {Kitagawa}\ \emph {et~al.}(2018)\citenamefont
		{Kitagawa}, \citenamefont {Takayama}, \citenamefont {Matsumoto},
		\citenamefont {Kato}, \citenamefont {Takano}, \citenamefont {Kishimoto},
		\citenamefont {Bette}, \citenamefont {Dinnebier}, \citenamefont {Jackeli},\
		and\ \citenamefont {Takagi}}]{Kitagawa2018}%
	\BibitemOpen
	\bibfield  {author} {\bibinfo {author} {\bibfnamefont {K.}~\bibnamefont
			{Kitagawa}}, \bibinfo {author} {\bibfnamefont {T.}~\bibnamefont {Takayama}},
		\bibinfo {author} {\bibfnamefont {Y.}~\bibnamefont {Matsumoto}}, \bibinfo
		{author} {\bibfnamefont {A.}~\bibnamefont {Kato}}, \bibinfo {author}
		{\bibfnamefont {R.}~\bibnamefont {Takano}}, \bibinfo {author} {\bibfnamefont
			{Y.}~\bibnamefont {Kishimoto}}, \bibinfo {author} {\bibfnamefont
			{S.}~\bibnamefont {Bette}}, \bibinfo {author} {\bibfnamefont
			{R.}~\bibnamefont {Dinnebier}}, \bibinfo {author} {\bibfnamefont
			{G.}~\bibnamefont {Jackeli}}, \ and\ \bibinfo {author} {\bibfnamefont
			{H.}~\bibnamefont {Takagi}},\ }\bibfield  {title} {\enquote {\bibinfo {title}
			{{A spin--orbital-entangled quantum liquid on a honeycomb lattice}},}\ }\href
	{http://dx.doi.org/10.1038/nature25482} {\bibfield  {journal} {\bibinfo
			{journal} {Nature}\ }\textbf {\bibinfo {volume} {554}},\ \bibinfo {pages}
		{341} (\bibinfo {year} {2018})}\BibitemShut {NoStop}%
	\bibitem [{\citenamefont {Plumb}\ \emph {et~al.}(2014)\citenamefont {Plumb},
		\citenamefont {Clancy}, \citenamefont {Sandilands}, \citenamefont {Shankar},
		\citenamefont {Hu}, \citenamefont {Burch}, \citenamefont {Kee},\ and\
		\citenamefont {Kim}}]{Plumb2014}%
	\BibitemOpen
	\bibfield  {author} {\bibinfo {author} {\bibfnamefont {K.~W.}\ \bibnamefont
			{Plumb}}, \bibinfo {author} {\bibfnamefont {J.~P.}\ \bibnamefont {Clancy}},
		\bibinfo {author} {\bibfnamefont {L.~J.}\ \bibnamefont {Sandilands}},
		\bibinfo {author} {\bibfnamefont {V.~Vijay}\ \bibnamefont {Shankar}},
		\bibinfo {author} {\bibfnamefont {Y.~F.}\ \bibnamefont {Hu}}, \bibinfo
		{author} {\bibfnamefont {K.~S.}\ \bibnamefont {Burch}}, \bibinfo {author}
		{\bibfnamefont {H.-Y.}\ \bibnamefont {Kee}}, \ and\ \bibinfo {author}
		{\bibfnamefont {Y.-J.}\ \bibnamefont {Kim}},\ }\bibfield  {title} {\enquote
		{\bibinfo {title} {{$\ensuremath{\alpha}$-RuCl$_{3}$: A spin-orbit assisted
					Mott insulator on a honeycomb lattice}},}\ }\href {\doibase
		10.1103/PhysRevB.90.041112} {\bibfield  {journal} {\bibinfo  {journal} {Phys.
				Rev. B}\ }\textbf {\bibinfo {volume} {90}},\ \bibinfo {pages} {041112}
		(\bibinfo {year} {2014})}\BibitemShut {NoStop}%
	\bibitem [{\citenamefont {Kasahara}\ \emph {et~al.}(2018)\citenamefont
		{Kasahara}, \citenamefont {Ohnishi}, \citenamefont {Mizukami}, \citenamefont
		{Tanaka}, \citenamefont {Ma}, \citenamefont {Sugii}, \citenamefont {Kurita},
		\citenamefont {Tanaka}, \citenamefont {Nasu}, \citenamefont {Motome},
		\citenamefont {Shibauchi},\ and\ \citenamefont {Matsuda}}]{Kasahara2018}%
	\BibitemOpen
	\bibfield  {author} {\bibinfo {author} {\bibfnamefont {Y.}~\bibnamefont
			{Kasahara}}, \bibinfo {author} {\bibfnamefont {T.}~\bibnamefont {Ohnishi}},
		\bibinfo {author} {\bibfnamefont {Y.}~\bibnamefont {Mizukami}}, \bibinfo
		{author} {\bibfnamefont {O.}~\bibnamefont {Tanaka}}, \bibinfo {author}
		{\bibfnamefont {S.}~\bibnamefont {Ma}}, \bibinfo {author} {\bibfnamefont
			{K.}~\bibnamefont {Sugii}}, \bibinfo {author} {\bibfnamefont
			{N.}~\bibnamefont {Kurita}}, \bibinfo {author} {\bibfnamefont
			{H.}~\bibnamefont {Tanaka}}, \bibinfo {author} {\bibfnamefont
			{J.}~\bibnamefont {Nasu}}, \bibinfo {author} {\bibfnamefont {Y.}~\bibnamefont
			{Motome}}, \bibinfo {author} {\bibfnamefont {T.}~\bibnamefont {Shibauchi}}, \
		and\ \bibinfo {author} {\bibfnamefont {Y.}~\bibnamefont {Matsuda}},\
	}\bibfield  {title} {\enquote {\bibinfo {title} {{Majorana quantization and
					half-integer thermal quantum Hall effect in a Kitaev spin liquid}},}\ }\href
	{\doibase 10.1038/s41586-018-0274-0} {\bibfield  {journal} {\bibinfo
			{journal} {Nature}\ }\textbf {\bibinfo {volume} {559}},\ \bibinfo {pages}
		{227} (\bibinfo {year} {2018})}\BibitemShut {NoStop}%
	\bibitem [{\citenamefont {Kitaev}(2006)}]{Kitaev2006}%
	\BibitemOpen
	\bibfield  {author} {\bibinfo {author} {\bibfnamefont {A.}~\bibnamefont
			{Kitaev}},\ }\bibfield  {title} {\enquote {\bibinfo {title} {{Anyons in an
					exactly solved model and beyond}},}\ }\href {\doibase
		http://dx.doi.org/10.1016/j.aop.2005.10.005} {\bibfield  {journal} {\bibinfo
			{journal} {Ann. Phys.}\ }\textbf {\bibinfo {volume} {321}},\ \bibinfo {pages}
		{2} (\bibinfo {year} {2006})}\BibitemShut {NoStop}%
	\bibitem [{\citenamefont {Wakeshima}\ \emph {et~al.}(2000)\citenamefont
		{Wakeshima}, \citenamefont {Harada},\ and\ \citenamefont
		{Hinatsu}}]{Wakeshima2000}%
	\BibitemOpen
	\bibfield  {author} {\bibinfo {author} {\bibfnamefont {M.}~\bibnamefont
			{Wakeshima}}, \bibinfo {author} {\bibfnamefont {D.}~\bibnamefont {Harada}}, \
		and\ \bibinfo {author} {\bibfnamefont {Y.}~\bibnamefont {Hinatsu}},\
	}\bibfield  {title} {\enquote {\bibinfo {title} {{Crystal structures and
					magnetic properties of ordered perovskites Ba$_2$LnIrO$_6$
					(Ln=lanthanide)}},}\ }\href {\doibase 10.1039/A907586K} {\bibfield  {journal}
		{\bibinfo  {journal} {J. Mater. Chem.}\ }\textbf {\bibinfo {volume} {10}},\
		\bibinfo {pages} {419} (\bibinfo {year} {2000})}\BibitemShut {NoStop}%
	\bibitem [{\citenamefont {Kockelmann}\ \emph {et~al.}(2006)\citenamefont
		{Kockelmann}, \citenamefont {Adroja}, \citenamefont {Hillier}, \citenamefont
		{Wakeshima}, \citenamefont {Izumiyama}, \citenamefont {Hinatsu},
		\citenamefont {Knight}, \citenamefont {Visser},\ and\ \citenamefont
		{Rainford}}]{Kockelmann06}%
	\BibitemOpen
	\bibfield  {author} {\bibinfo {author} {\bibfnamefont {W.}~\bibnamefont
			{Kockelmann}}, \bibinfo {author} {\bibfnamefont {D.T.}\ \bibnamefont
			{Adroja}}, \bibinfo {author} {\bibfnamefont {A.D.}\ \bibnamefont {Hillier}},
		\bibinfo {author} {\bibfnamefont {M.}~\bibnamefont {Wakeshima}}, \bibinfo
		{author} {\bibfnamefont {Y.}~\bibnamefont {Izumiyama}}, \bibinfo {author}
		{\bibfnamefont {Y.}~\bibnamefont {Hinatsu}}, \bibinfo {author} {\bibfnamefont
			{K.S.}\ \bibnamefont {Knight}}, \bibinfo {author} {\bibfnamefont
			{D.}~\bibnamefont {Visser}}, \ and\ \bibinfo {author} {\bibfnamefont {B.D.}\
			\bibnamefont {Rainford}},\ }\bibfield  {title} {\enquote {\bibinfo {title}
			{{Neutron diffraction and inelastic neutron scattering investigations of the
					ordered double perovskite Ba$_2$PrIrO$_6$}},}\ }\href@noop {} {\bibfield
		{journal} {\bibinfo  {journal} {Physica B}\ }\textbf {\bibinfo {volume}
			{378}},\ \bibinfo {pages} {543} (\bibinfo {year} {2006})}\BibitemShut
	{NoStop}%
	\bibitem [{\citenamefont {Nelmes}()}]{Nelmes1987}%
	\BibitemOpen
	\bibfield  {author} {\bibinfo {author} {\bibfnamefont {R.~J.}\ \bibnamefont
			{Nelmes}},\ }\bibfield  {title} {\enquote {\bibinfo {title} {{Structural
					studies of KDP and the KDP-type transition by neutron and x-ray diffraction:
					1970-1985}},}\ }\href@noop {} {\bibfield  {journal} {\bibinfo  {journal}
			{Ferroelectrics}\ }\textbf {\bibinfo {volume} {71}},\ \bibinfo {pages}
		{87}}\BibitemShut {NoStop}%
	\bibitem [{\citenamefont {Braden}\ \emph {et~al.}(2001)\citenamefont {Braden},
		\citenamefont {Meven}, \citenamefont {Reichardt}, \citenamefont
		{Pintschovius}, \citenamefont {Fernandez-Diaz}, \citenamefont {Heger},
		\citenamefont {Nakamura},\ and\ \citenamefont {Fujita}}]{Braden2001}%
	\BibitemOpen
	\bibfield  {author} {\bibinfo {author} {\bibfnamefont {M.}~\bibnamefont
			{Braden}}, \bibinfo {author} {\bibfnamefont {M.}~\bibnamefont {Meven}},
		\bibinfo {author} {\bibfnamefont {W.}~\bibnamefont {Reichardt}}, \bibinfo
		{author} {\bibfnamefont {L.}~\bibnamefont {Pintschovius}}, \bibinfo {author}
		{\bibfnamefont {M.~T.}\ \bibnamefont {Fernandez-Diaz}}, \bibinfo {author}
		{\bibfnamefont {G.}~\bibnamefont {Heger}}, \bibinfo {author} {\bibfnamefont
			{F.}~\bibnamefont {Nakamura}}, \ and\ \bibinfo {author} {\bibfnamefont
			{T.}~\bibnamefont {Fujita}},\ }\bibfield  {title} {\enquote {\bibinfo {title}
			{{Analysis of the local structure by single-crystal neutron scattering in
					${\mathrm{La}}_{1.85}{\mathrm{Sr}}_{0.15}{\mathrm{CuO}}_{4}$}},}\ }\href
	{\doibase 10.1103/PhysRevB.63.140510} {\bibfield  {journal} {\bibinfo
			{journal} {Phys. Rev. B}\ }\textbf {\bibinfo {volume} {63}},\ \bibinfo
		{pages} {140510} (\bibinfo {year} {2001})}\BibitemShut {NoStop}%
	\bibitem [{\citenamefont {Vasala}\ and\ \citenamefont
		{Karppinen}(2015)}]{Vasala2015}%
	\BibitemOpen
	\bibfield  {author} {\bibinfo {author} {\bibfnamefont {S.}~\bibnamefont
			{Vasala}}\ and\ \bibinfo {author} {\bibfnamefont {M.}~\bibnamefont
			{Karppinen}},\ }\bibfield  {title} {\enquote {\bibinfo {title}
			{{$A_2B$'$B$''O$_6$ perovskites: A review}},}\ }\href@noop {} {\bibfield
		{journal} {\bibinfo  {journal} {Prog. Solid State Chem.}\ }\textbf {\bibinfo
			{volume} {43}},\ \bibinfo {pages} {1} (\bibinfo {year} {2015})}\BibitemShut
	{NoStop}%
	\bibitem [{Note1()}]{Note1}%
	\BibitemOpen
	\bibinfo {note} {$\chi _{dia}\simeq -1.7\cdot 10^{-4}$\protect \tmspace
		+\thinmuskip {.1667em}emu/mol results from the tabulated values \cite
		{Bain2008} for the ions in Ba$_2$CeIrO$_6$.}\BibitemShut {Stop}%
	\bibitem [{\citenamefont {Moretti~Sala}\ \emph {et~al.}(2014)\citenamefont
		{Moretti~Sala}, \citenamefont {Boseggia}, \citenamefont {McMorrow},\ and\
		\citenamefont {Monaco}}]{Moretti_PRL2014}%
	\BibitemOpen
	\bibfield  {author} {\bibinfo {author} {\bibfnamefont {M.}~\bibnamefont
			{Moretti~Sala}}, \bibinfo {author} {\bibfnamefont {S.}~\bibnamefont
			{Boseggia}}, \bibinfo {author} {\bibfnamefont {D.~F.}\ \bibnamefont
			{McMorrow}}, \ and\ \bibinfo {author} {\bibfnamefont {G.}~\bibnamefont
			{Monaco}},\ }\bibfield  {title} {\enquote {\bibinfo {title} {{Resonant X-Ray
					Scattering and the ${j}_{\mathrm{eff}}\mathbf{=}1/2$ Electronic Ground State
					in Iridate Perovskites}},}\ }\href {\doibase 10.1103/PhysRevLett.112.026403}
	{\bibfield  {journal} {\bibinfo  {journal} {Phys. Rev. Lett.}\ }\textbf
		{\bibinfo {volume} {112}},\ \bibinfo {pages} {026403} (\bibinfo {year}
		{2014})}\BibitemShut {NoStop}%
	\bibitem [{\citenamefont {Gretarsson}\ \emph {et~al.}(2013)\citenamefont
		{Gretarsson}, \citenamefont {Clancy}, \citenamefont {Liu}, \citenamefont
		{Hill}, \citenamefont {Bozin}, \citenamefont {Singh}, \citenamefont {Manni},
		\citenamefont {Gegenwart}, \citenamefont {Kim}, \citenamefont {Said},
		\citenamefont {Casa}, \citenamefont {Gog}, \citenamefont {Upton},
		\citenamefont {Kim}, \citenamefont {Yu}, \citenamefont {Katukuri},
		\citenamefont {Hozoi}, \citenamefont {van~den Brink},\ and\ \citenamefont
		{Kim}}]{Gretarsson13}%
	\BibitemOpen
	\bibfield  {author} {\bibinfo {author} {\bibfnamefont {H.}~\bibnamefont
			{Gretarsson}}, \bibinfo {author} {\bibfnamefont {J.~P.}\ \bibnamefont
			{Clancy}}, \bibinfo {author} {\bibfnamefont {X.}~\bibnamefont {Liu}},
		\bibinfo {author} {\bibfnamefont {J.~P.}\ \bibnamefont {Hill}}, \bibinfo
		{author} {\bibfnamefont {E.}~\bibnamefont {Bozin}}, \bibinfo {author}
		{\bibfnamefont {Y.}~\bibnamefont {Singh}}, \bibinfo {author} {\bibfnamefont
			{S.}~\bibnamefont {Manni}}, \bibinfo {author} {\bibfnamefont
			{P.}~\bibnamefont {Gegenwart}}, \bibinfo {author} {\bibfnamefont
			{J.}~\bibnamefont {Kim}}, \bibinfo {author} {\bibfnamefont {A.~H.}\
			\bibnamefont {Said}}, \bibinfo {author} {\bibfnamefont {D.}~\bibnamefont
			{Casa}}, \bibinfo {author} {\bibfnamefont {T.}~\bibnamefont {Gog}}, \bibinfo
		{author} {\bibfnamefont {M.~H.}\ \bibnamefont {Upton}}, \bibinfo {author}
		{\bibfnamefont {H.-S.}\ \bibnamefont {Kim}}, \bibinfo {author} {\bibfnamefont
			{J.}~\bibnamefont {Yu}}, \bibinfo {author} {\bibfnamefont {V.~M.}\
			\bibnamefont {Katukuri}}, \bibinfo {author} {\bibfnamefont {L.}~\bibnamefont
			{Hozoi}}, \bibinfo {author} {\bibfnamefont {J.}~\bibnamefont {van~den
				Brink}}, \ and\ \bibinfo {author} {\bibfnamefont {Y.-J.}\ \bibnamefont
			{Kim}},\ }\bibfield  {title} {\enquote {\bibinfo {title} {{Crystal-Field
					Splitting and Correlation Effect on the Electronic Structure of
					${A}_{2}{\mathrm{IrO}}_{3}$}},}\ }\href {\doibase
		10.1103/PhysRevLett.110.076402} {\bibfield  {journal} {\bibinfo  {journal}
			{Phys. Rev. Lett.}\ }\textbf {\bibinfo {volume} {110}},\ \bibinfo {pages}
		{076402} (\bibinfo {year} {2013})}\BibitemShut {NoStop}%
	\bibitem [{\citenamefont {Kim}\ \emph {et~al.}(2014{\natexlab{b}})\citenamefont
		{Kim}, \citenamefont {Daghofer}, \citenamefont {Said}, \citenamefont {Gog},
		\citenamefont {van~den Brink}, \citenamefont {Khaliullin},\ and\
		\citenamefont {Kim}}]{KimRIXS14}%
	\BibitemOpen
	\bibfield  {author} {\bibinfo {author} {\bibfnamefont {J.}~\bibnamefont
			{Kim}}, \bibinfo {author} {\bibfnamefont {M.}~\bibnamefont {Daghofer}},
		\bibinfo {author} {\bibfnamefont {A.~H.}\ \bibnamefont {Said}}, \bibinfo
		{author} {\bibfnamefont {T.}~\bibnamefont {Gog}}, \bibinfo {author}
		{\bibfnamefont {J.}~\bibnamefont {van~den Brink}}, \bibinfo {author}
		{\bibfnamefont {G.}~\bibnamefont {Khaliullin}}, \ and\ \bibinfo {author}
		{\bibfnamefont {B.~J.}\ \bibnamefont {Kim}},\ }\bibfield  {title} {\enquote
		{\bibinfo {title} {{Excitonic quasiparticles in a spin-orbit Mott
					insulator}},}\ }\href {\doibase 10.1038/ncomms5453} {\bibfield  {journal}
		{\bibinfo  {journal} {Nat. Commun.}\ }\textbf {\bibinfo {volume} {5}},\
		\bibinfo {pages} {4453} (\bibinfo {year} {2014}{\natexlab{b}})}\BibitemShut
	{NoStop}%
	\bibitem [{\citenamefont {Rossi}\ \emph {et~al.}(2017)\citenamefont {Rossi},
		\citenamefont {Retegan}, \citenamefont {Giacobbe}, \citenamefont {Fumagalli},
		\citenamefont {Efimenko}, \citenamefont {Kulka}, \citenamefont {Wohlfeld},
		\citenamefont {Gubanov},\ and\ \citenamefont {Moretti~Sala}}]{Rossi17}%
	\BibitemOpen
	\bibfield  {author} {\bibinfo {author} {\bibfnamefont {M.}~\bibnamefont
			{Rossi}}, \bibinfo {author} {\bibfnamefont {M.}~\bibnamefont {Retegan}},
		\bibinfo {author} {\bibfnamefont {C.}~\bibnamefont {Giacobbe}}, \bibinfo
		{author} {\bibfnamefont {R.}~\bibnamefont {Fumagalli}}, \bibinfo {author}
		{\bibfnamefont {A.}~\bibnamefont {Efimenko}}, \bibinfo {author}
		{\bibfnamefont {T.}~\bibnamefont {Kulka}}, \bibinfo {author} {\bibfnamefont
			{K.}~\bibnamefont {Wohlfeld}}, \bibinfo {author} {\bibfnamefont {A.~I.}\
			\bibnamefont {Gubanov}}, \ and\ \bibinfo {author} {\bibfnamefont
			{M.}~\bibnamefont {Moretti~Sala}},\ }\bibfield  {title} {\enquote {\bibinfo
			{title} {{Possibility to realize spin-orbit-induced correlated physics in
					iridium fluorides}},}\ }\href {\doibase 10.1103/PhysRevB.95.235161}
	{\bibfield  {journal} {\bibinfo  {journal} {Phys. Rev. B}\ }\textbf {\bibinfo
			{volume} {95}},\ \bibinfo {pages} {235161} (\bibinfo {year}
		{2017})}\BibitemShut {NoStop}%
	\bibitem [{\citenamefont {Liu}\ \emph {et~al.}(2012)\citenamefont {Liu},
		\citenamefont {Katukuri}, \citenamefont {Hozoi}, \citenamefont {Yin},
		\citenamefont {Dean}, \citenamefont {Upton}, \citenamefont {Kim},
		\citenamefont {Casa}, \citenamefont {Said}, \citenamefont {Gog},
		\citenamefont {Qi}, \citenamefont {Cao}, \citenamefont {Tsvelik},
		\citenamefont {van~den Brink},\ and\ \citenamefont {Hill}}]{Liu12}%
	\BibitemOpen
	\bibfield  {author} {\bibinfo {author} {\bibfnamefont {X.}~\bibnamefont
			{Liu}}, \bibinfo {author} {\bibfnamefont {V.~M.}\ \bibnamefont {Katukuri}},
		\bibinfo {author} {\bibfnamefont {L.}~\bibnamefont {Hozoi}}, \bibinfo
		{author} {\bibfnamefont {W.-G.}\ \bibnamefont {Yin}}, \bibinfo {author}
		{\bibfnamefont {M.~P.~M.}\ \bibnamefont {Dean}}, \bibinfo {author}
		{\bibfnamefont {M.~H.}\ \bibnamefont {Upton}}, \bibinfo {author}
		{\bibfnamefont {J.}~\bibnamefont {Kim}}, \bibinfo {author} {\bibfnamefont
			{D.}~\bibnamefont {Casa}}, \bibinfo {author} {\bibfnamefont {A.}~\bibnamefont
			{Said}}, \bibinfo {author} {\bibfnamefont {T.}~\bibnamefont {Gog}}, \bibinfo
		{author} {\bibfnamefont {T.~F.}\ \bibnamefont {Qi}}, \bibinfo {author}
		{\bibfnamefont {G.}~\bibnamefont {Cao}}, \bibinfo {author} {\bibfnamefont
			{A.~M.}\ \bibnamefont {Tsvelik}}, \bibinfo {author} {\bibfnamefont
			{J.}~\bibnamefont {van~den Brink}}, \ and\ \bibinfo {author} {\bibfnamefont
			{J.~P.}\ \bibnamefont {Hill}},\ }\bibfield  {title} {\enquote {\bibinfo
			{title} {{Testing the Validity of the Strong Spin-Orbit-Coupling Limit for
					Octahedrally Coordinated Iridate Compounds in a Model System
					${\mathrm{Sr}}_{3}{\mathrm{CuIrO}}_{6}$}},}\ }\href {\doibase
		10.1103/PhysRevLett.109.157401} {\bibfield  {journal} {\bibinfo  {journal}
			{Phys. Rev. Lett.}\ }\textbf {\bibinfo {volume} {109}},\ \bibinfo {pages}
		{157401} (\bibinfo {year} {2012})}\BibitemShut {NoStop}%
	\bibitem [{\citenamefont {Sala}\ \emph {et~al.}(2014)\citenamefont {Sala},
		\citenamefont {Ohgushi}, \citenamefont {Al-Zein}, \citenamefont {Hirata},
		\citenamefont {Monaco},\ and\ \citenamefont {Krisch}}]{Moretti14}%
	\BibitemOpen
	\bibfield  {author} {\bibinfo {author} {\bibfnamefont {M.~Moretti}\
			\bibnamefont {Sala}}, \bibinfo {author} {\bibfnamefont {K.}~\bibnamefont
			{Ohgushi}}, \bibinfo {author} {\bibfnamefont {A.}~\bibnamefont {Al-Zein}},
		\bibinfo {author} {\bibfnamefont {Y.}~\bibnamefont {Hirata}}, \bibinfo
		{author} {\bibfnamefont {G.}~\bibnamefont {Monaco}}, \ and\ \bibinfo {author}
		{\bibfnamefont {M.}~\bibnamefont {Krisch}},\ }\bibfield  {title} {\enquote
		{\bibinfo {title} {{CaIrO$_3$: A Spin-Orbit Mott Insulator Beyond the
					${j}_{\text{eff}}$\,=\,$1/2$ Ground State}},}\ }\href {\doibase
		10.1103/PhysRevLett.112.176402} {\bibfield  {journal} {\bibinfo  {journal}
			{Phys. Rev. Lett.}\ }\textbf {\bibinfo {volume} {112}},\ \bibinfo {pages}
		{176402} (\bibinfo {year} {2014})}\BibitemShut {NoStop}%
	\bibitem [{\citenamefont {Crawford}\ \emph {et~al.}(1994)\citenamefont
		{Crawford}, \citenamefont {Subramanian}, \citenamefont {Harlow},
		\citenamefont {Fernandez-Baca}, \citenamefont {Wang},\ and\ \citenamefont
		{Johnston}}]{Crawford94}%
	\BibitemOpen
	\bibfield  {author} {\bibinfo {author} {\bibfnamefont {M.~K.}\ \bibnamefont
			{Crawford}}, \bibinfo {author} {\bibfnamefont {M.~A.}\ \bibnamefont
			{Subramanian}}, \bibinfo {author} {\bibfnamefont {R.~L.}\ \bibnamefont
			{Harlow}}, \bibinfo {author} {\bibfnamefont {J.~A.}\ \bibnamefont
			{Fernandez-Baca}}, \bibinfo {author} {\bibfnamefont {Z.~R.}\ \bibnamefont
			{Wang}}, \ and\ \bibinfo {author} {\bibfnamefont {D.~C.}\ \bibnamefont
			{Johnston}},\ }\bibfield  {title} {\enquote {\bibinfo {title} {{Structural
					and magnetic studies of ${\mathrm{Sr}}_{2}$${\mathrm{IrO}}_{4}$}},}\ }\href
	{\doibase 10.1103/PhysRevB.49.9198} {\bibfield  {journal} {\bibinfo
			{journal} {Phys. Rev. B}\ }\textbf {\bibinfo {volume} {49}},\ \bibinfo
		{pages} {9198} (\bibinfo {year} {1994})}\BibitemShut {NoStop}%
	\bibitem [{\citenamefont {Moretti~Sala}\ \emph {et~al.}(2013)\citenamefont
		{Moretti~Sala}, \citenamefont {Henriquet}, \citenamefont {Simonelli},
		\citenamefont {Verbeni},\ and\ \citenamefont {Monaco}}]{Moretti13}%
	\BibitemOpen
	\bibfield  {author} {\bibinfo {author} {\bibfnamefont {M.}~\bibnamefont
			{Moretti~Sala}}, \bibinfo {author} {\bibfnamefont {C.}~\bibnamefont
			{Henriquet}}, \bibinfo {author} {\bibfnamefont {L.}~\bibnamefont
			{Simonelli}}, \bibinfo {author} {\bibfnamefont {R.}~\bibnamefont {Verbeni}},
		\ and\ \bibinfo {author} {\bibfnamefont {G.}~\bibnamefont {Monaco}},\
	}\bibfield  {title} {\enquote {\bibinfo {title} {{High energy-resolution
					set-up for Ir $L_3$ edge RIXS experiments}},}\ }\href@noop {} {\bibfield
		{journal} {\bibinfo  {journal} {J. Electron Spectrosc. Relat. Phenom.}\
		}\textbf {\bibinfo {volume} {188}},\ \bibinfo {pages} {150} (\bibinfo {year}
		{2013})}\BibitemShut {NoStop}%
	\bibitem [{\citenamefont {Moretti~Sala}\ \emph {et~al.}(2018)\citenamefont
		{Moretti~Sala}, \citenamefont {Martel}, \citenamefont {Henriquet},
		\citenamefont {Al~Zein}, \citenamefont {Simonelli}, \citenamefont {Sahle},
		\citenamefont {Gonzalez}, \citenamefont {Lagier}, \citenamefont {Ponchut},
		\citenamefont {Huotari}, \citenamefont {Verbeni}, \citenamefont {Krisch},\
		and\ \citenamefont {Monaco}}]{Moretti18}%
	\BibitemOpen
	\bibfield  {author} {\bibinfo {author} {\bibfnamefont {M.}~\bibnamefont
			{Moretti~Sala}}, \bibinfo {author} {\bibfnamefont {K.}~\bibnamefont
			{Martel}}, \bibinfo {author} {\bibfnamefont {C.}~\bibnamefont {Henriquet}},
		\bibinfo {author} {\bibfnamefont {A.}~\bibnamefont {Al~Zein}}, \bibinfo
		{author} {\bibfnamefont {L.}~\bibnamefont {Simonelli}}, \bibinfo {author}
		{\bibfnamefont {Ch.J.}\ \bibnamefont {Sahle}}, \bibinfo {author}
		{\bibfnamefont {H.}~\bibnamefont {Gonzalez}}, \bibinfo {author}
		{\bibfnamefont {M.-C.}\ \bibnamefont {Lagier}}, \bibinfo {author}
		{\bibfnamefont {C.}~\bibnamefont {Ponchut}}, \bibinfo {author} {\bibfnamefont
			{S.}~\bibnamefont {Huotari}}, \bibinfo {author} {\bibfnamefont
			{R.}~\bibnamefont {Verbeni}}, \bibinfo {author} {\bibfnamefont
			{M.}~\bibnamefont {Krisch}}, \ and\ \bibinfo {author} {\bibfnamefont
			{G.}~\bibnamefont {Monaco}},\ }\bibfield  {title} {\enquote {\bibinfo {title}
			{{A high-energy-resolution resonant inelastic X-ray scattering spectrometer
					at ID20 of the European Synchrotron Radiation Facility}},}\ }\href@noop {}
	{\bibfield  {journal} {\bibinfo  {journal} {J. Synchrotron Rad.}\ }\textbf
		{\bibinfo {volume} {25}},\ \bibinfo {pages} {580} (\bibinfo {year}
		{2018})}\BibitemShut {NoStop}%
	\bibitem [{\citenamefont {Revelli}\ \emph {et~al.}(2019)\citenamefont
		{Revelli}, \citenamefont {Moretti~Sala}, \citenamefont {Monaco},
		\citenamefont {Becker}, \citenamefont {Bohat\'{y}}, \citenamefont {Hermanns},
		\citenamefont {Koethe}, \citenamefont {Fr\"{o}hlich}, \citenamefont
		{Warzanowski}, \citenamefont {Lorenz}, \citenamefont {Streltsov},
		\citenamefont {van Loosdrecht}, \citenamefont {Khomskii}, \citenamefont
		{van~den Brink},\ and\ \citenamefont {Gr\"{u}ninger}}]{Revelli2019}%
	\BibitemOpen
	\bibfield  {author} {\bibinfo {author} {\bibfnamefont {A.}~\bibnamefont
			{Revelli}}, \bibinfo {author} {\bibfnamefont {M.}~\bibnamefont
			{Moretti~Sala}}, \bibinfo {author} {\bibfnamefont {G.}~\bibnamefont
			{Monaco}}, \bibinfo {author} {\bibfnamefont {P.}~\bibnamefont {Becker}},
		\bibinfo {author} {\bibfnamefont {L.}~\bibnamefont {Bohat\'{y}}}, \bibinfo
		{author} {\bibfnamefont {M.}~\bibnamefont {Hermanns}}, \bibinfo {author}
		{\bibfnamefont {T.C.}\ \bibnamefont {Koethe}}, \bibinfo {author}
		{\bibfnamefont {T.}~\bibnamefont {Fr\"{o}hlich}}, \bibinfo {author}
		{\bibfnamefont {P.}~\bibnamefont {Warzanowski}}, \bibinfo {author}
		{\bibfnamefont {T.}~\bibnamefont {Lorenz}}, \bibinfo {author} {\bibfnamefont
			{S.V.}\ \bibnamefont {Streltsov}}, \bibinfo {author} {\bibfnamefont {P.H.M.}\
			\bibnamefont {van Loosdrecht}}, \bibinfo {author} {\bibfnamefont {D.I.}\
			\bibnamefont {Khomskii}}, \bibinfo {author} {\bibfnamefont {J.}~\bibnamefont
			{van~den Brink}}, \ and\ \bibinfo {author} {\bibfnamefont {M.}~\bibnamefont
			{Gr\"{u}ninger}},\ }\bibfield  {title} {\enquote {\bibinfo {title} {Resonant
				inelastic x-ray incarnation of young's double-slit experiment},}\ }\href@noop
	{} {\bibfield  {journal} {\bibinfo  {journal} {Sci. Adv.}\ }\textbf {\bibinfo
			{volume} {5}},\ \bibinfo {pages} {eaav4020} (\bibinfo {year}
		{2019})}\BibitemShut {NoStop}%
	\bibitem [{\citenamefont {Wang}\ and\ \citenamefont
		{Zhou}(2005)}]{Wang05Pearson}%
	\BibitemOpen
	\bibfield  {author} {\bibinfo {author} {\bibfnamefont {H.}~\bibnamefont
			{Wang}}\ and\ \bibinfo {author} {\bibfnamefont {J.}~\bibnamefont {Zhou}},\
	}\bibfield  {title} {\enquote {\bibinfo {title} {{{Numerical conversion
						between the Pearson VII and pseudo-Voigt functions}}},}\ }\href@noop {}
	{\bibfield  {journal} {\bibinfo  {journal} {J. Appl. Cryst.}\ }\textbf
		{\bibinfo {volume} {38}},\ \bibinfo {pages} {830} (\bibinfo {year}
		{2005})}\BibitemShut {NoStop}%
	\bibitem [{\citenamefont {Kim}\ \emph {et~al.}(2012)\citenamefont {Kim},
		\citenamefont {Casa}, \citenamefont {Upton}, \citenamefont {Gog},
		\citenamefont {Kim}, \citenamefont {Mitchell}, \citenamefont {van
			Veenendaal}, \citenamefont {Daghofer}, \citenamefont {van~den Brink},
		\citenamefont {Khaliullin},\ and\ \citenamefont {Kim}}]{KimRIXS12}%
	\BibitemOpen
	\bibfield  {author} {\bibinfo {author} {\bibfnamefont {J.}~\bibnamefont
			{Kim}}, \bibinfo {author} {\bibfnamefont {D.}~\bibnamefont {Casa}}, \bibinfo
		{author} {\bibfnamefont {M.~H.}\ \bibnamefont {Upton}}, \bibinfo {author}
		{\bibfnamefont {T.}~\bibnamefont {Gog}}, \bibinfo {author} {\bibfnamefont
			{Y.-J.}\ \bibnamefont {Kim}}, \bibinfo {author} {\bibfnamefont {J.~F.}\
			\bibnamefont {Mitchell}}, \bibinfo {author} {\bibfnamefont {M.}~\bibnamefont
			{van Veenendaal}}, \bibinfo {author} {\bibfnamefont {M.}~\bibnamefont
			{Daghofer}}, \bibinfo {author} {\bibfnamefont {J.}~\bibnamefont {van~den
				Brink}}, \bibinfo {author} {\bibfnamefont {G.}~\bibnamefont {Khaliullin}}, \
		and\ \bibinfo {author} {\bibfnamefont {B.~J.}\ \bibnamefont {Kim}},\
	}\bibfield  {title} {\enquote {\bibinfo {title} {{Magnetic Excitation Spectra
					of ${\mathrm{Sr}}_{2}{\mathrm{IrO}}_{4}$ Probed by Resonant Inelastic X-Ray
					Scattering: Establishing Links to Cuprate Superconductors}},}\ }\href
	{\doibase 10.1103/PhysRevLett.108.177003} {\bibfield  {journal} {\bibinfo
			{journal} {Phys. Rev. Lett.}\ }\textbf {\bibinfo {volume} {108}},\ \bibinfo
		{pages} {177003} (\bibinfo {year} {2012})}\BibitemShut {NoStop}%
	\bibitem [{\citenamefont {Cook}\ \emph {et~al.}(2015)\citenamefont {Cook},
		\citenamefont {Matern}, \citenamefont {Hickey}, \citenamefont {Aczel},\ and\
		\citenamefont {Paramekanti}}]{Cook2015}%
	\BibitemOpen
	\bibfield  {author} {\bibinfo {author} {\bibfnamefont {A.~M.}\ \bibnamefont
			{Cook}}, \bibinfo {author} {\bibfnamefont {S.}~\bibnamefont {Matern}},
		\bibinfo {author} {\bibfnamefont {C.}~\bibnamefont {Hickey}}, \bibinfo
		{author} {\bibfnamefont {A.~A.}\ \bibnamefont {Aczel}}, \ and\ \bibinfo
		{author} {\bibfnamefont {A.}~\bibnamefont {Paramekanti}},\ }\bibfield
	{title} {\enquote {\bibinfo {title} {{Spin-orbit coupled
					${j}_{\mathrm{eff}}$\,=\,$1/2$ iridium moments on the geometrically
					frustrated fcc lattice}},}\ }\href {\doibase 10.1103/PhysRevB.92.020417}
	{\bibfield  {journal} {\bibinfo  {journal} {Phys. Rev. B}\ }\textbf {\bibinfo
			{volume} {92}},\ \bibinfo {pages} {020417} (\bibinfo {year}
		{2015})}\BibitemShut {NoStop}%
	\bibitem [{\citenamefont {Aczel}\ \emph {et~al.}(2016)\citenamefont {Aczel},
		\citenamefont {Cook}, \citenamefont {Williams}, \citenamefont {Calder},
		\citenamefont {Christianson}, \citenamefont {Cao}, \citenamefont {Mandrus},
		\citenamefont {Kim},\ and\ \citenamefont {Paramekanti}}]{Aczel2016}%
	\BibitemOpen
	\bibfield  {author} {\bibinfo {author} {\bibfnamefont {A.~A.}\ \bibnamefont
			{Aczel}}, \bibinfo {author} {\bibfnamefont {A.~M.}\ \bibnamefont {Cook}},
		\bibinfo {author} {\bibfnamefont {T.~J.}\ \bibnamefont {Williams}}, \bibinfo
		{author} {\bibfnamefont {S.}~\bibnamefont {Calder}}, \bibinfo {author}
		{\bibfnamefont {A.~D.}\ \bibnamefont {Christianson}}, \bibinfo {author}
		{\bibfnamefont {G.-X.}\ \bibnamefont {Cao}}, \bibinfo {author} {\bibfnamefont
			{D.}~\bibnamefont {Mandrus}}, \bibinfo {author} {\bibfnamefont {Y.-B.}\
			\bibnamefont {Kim}}, \ and\ \bibinfo {author} {\bibfnamefont
			{A.}~\bibnamefont {Paramekanti}},\ }\bibfield  {title} {\enquote {\bibinfo
			{title} {{Highly anisotropic exchange interactions of $j_{\rm eff}$\,=\,$1/2$
					iridium moments on the fcc lattice in ${\rm La}_{2}B{\rm IrO}_{6} (B={\rm
						Mg},{\rm Zn})$}},}\ }\href {\doibase 10.1103/PhysRevB.93.214426} {\bibfield
		{journal} {\bibinfo  {journal} {Phys. Rev. B}\ }\textbf {\bibinfo {volume}
			{93}},\ \bibinfo {pages} {214426} (\bibinfo {year} {2016})}\BibitemShut
	{NoStop}%
	\bibitem [{\citenamefont {Li}\ \emph {et~al.}(2017)\citenamefont {Li},
		\citenamefont {Li}, \citenamefont {Yu}, \citenamefont {Paramekanti},\ and\
		\citenamefont {Chen}}]{Gang2017}%
	\BibitemOpen
	\bibfield  {author} {\bibinfo {author} {\bibfnamefont {F.-Y.}\ \bibnamefont
			{Li}}, \bibinfo {author} {\bibfnamefont {Y.-D.}\ \bibnamefont {Li}}, \bibinfo
		{author} {\bibfnamefont {Y.}~\bibnamefont {Yu}}, \bibinfo {author}
		{\bibfnamefont {A.}~\bibnamefont {Paramekanti}}, \ and\ \bibinfo {author}
		{\bibfnamefont {G.}~\bibnamefont {Chen}},\ }\bibfield  {title} {\enquote
		{\bibinfo {title} {{Kitaev materials beyond iridates: Order by quantum
					disorder and Weyl magnons in rare-earth double perovskites}},}\ }\href
	{\doibase 10.1103/PhysRevB.95.085132} {\bibfield  {journal} {\bibinfo
			{journal} {Phys. Rev. B}\ }\textbf {\bibinfo {volume} {95}},\ \bibinfo
		{pages} {085132} (\bibinfo {year} {2017})}\BibitemShut {NoStop}%
	\bibitem [{\citenamefont {Winter}\ \emph {et~al.}(2016)\citenamefont {Winter},
		\citenamefont {Li}, \citenamefont {Jeschke},\ and\ \citenamefont
		{Valent\'{\i}}}]{Winter2016}%
	\BibitemOpen
	\bibfield  {author} {\bibinfo {author} {\bibfnamefont {S.~M.}\ \bibnamefont
			{Winter}}, \bibinfo {author} {\bibfnamefont {Y.}~\bibnamefont {Li}}, \bibinfo
		{author} {\bibfnamefont {H.~O.}\ \bibnamefont {Jeschke}}, \ and\ \bibinfo
		{author} {\bibfnamefont {R.}~\bibnamefont {Valent\'{\i}}},\ }\bibfield
	{title} {\enquote {\bibinfo {title} {{Challenges in design of Kitaev
					materials: Magnetic interactions from competing energy scales}},}\ }\href
	{\doibase 10.1103/PhysRevB.93.214431} {\bibfield  {journal} {\bibinfo
			{journal} {Phys. Rev. B}\ }\textbf {\bibinfo {volume} {93}},\ \bibinfo
		{pages} {214431} (\bibinfo {year} {2016})}\BibitemShut {NoStop}%
	\bibitem [{\citenamefont {Reuther}\ and\ \citenamefont
		{W{\"{o}}lfle}(2010)}]{Reuther2010}%
	\BibitemOpen
	\bibfield  {author} {\bibinfo {author} {\bibfnamefont {J.}~\bibnamefont
			{Reuther}}\ and\ \bibinfo {author} {\bibfnamefont {P.}~\bibnamefont
			{W{\"{o}}lfle}},\ }\bibfield  {title} {\enquote {\bibinfo {title}
			{{J$_1$-J$_2$ frustrated two-dimensional Heisenberg model: Random phase
					approximation and functional renormalization group}},}\ }\href {\doibase
		10.1103/PhysRevB.81.144410} {\bibfield  {journal} {\bibinfo  {journal} {Phys.
				Rev. B}\ }\textbf {\bibinfo {volume} {81}},\ \bibinfo {pages} {144410}
		(\bibinfo {year} {2010})}\BibitemShut {NoStop}%
	\bibitem [{\citenamefont {Baez}\ and\ \citenamefont
		{Reuther}(2017)}]{Baez2016}%
	\BibitemOpen
	\bibfield  {author} {\bibinfo {author} {\bibfnamefont {M.~L.}\ \bibnamefont
			{Baez}}\ and\ \bibinfo {author} {\bibfnamefont {J.}~\bibnamefont {Reuther}},\
	}\bibfield  {title} {\enquote {\bibinfo {title} {{Numerical treatment of spin
					systems with unrestricted spin length $S$: A functional renormalization group
					study}},}\ }\href {\doibase 10.1103/PhysRevB.96.045144} {\bibfield  {journal}
		{\bibinfo  {journal} {Phys. Rev. B}\ }\textbf {\bibinfo {volume} {96}},\
		\bibinfo {pages} {045144} (\bibinfo {year} {2017})}\BibitemShut {NoStop}%
	\bibitem [{\citenamefont {Buessen}\ \emph
		{et~al.}(2018{\natexlab{a}})\citenamefont {Buessen}, \citenamefont {Roscher},
		\citenamefont {Diehl},\ and\ \citenamefont {Trebst}}]{Buessen2018}%
	\BibitemOpen
	\bibfield  {author} {\bibinfo {author} {\bibfnamefont {F.~L.}\ \bibnamefont
			{Buessen}}, \bibinfo {author} {\bibfnamefont {D.}~\bibnamefont {Roscher}},
		\bibinfo {author} {\bibfnamefont {S.}~\bibnamefont {Diehl}}, \ and\ \bibinfo
		{author} {\bibfnamefont {S.}~\bibnamefont {Trebst}},\ }\bibfield  {title}
	{\enquote {\bibinfo {title} {{Functional renormalization group approach to
					$\text{SU}(N)$ Heisenberg models: Real-space renormalization group at
					arbitrary $N$}},}\ }\href {\doibase 10.1103/PhysRevB.97.064415} {\bibfield
		{journal} {\bibinfo  {journal} {Phys. Rev. B}\ }\textbf {\bibinfo {volume}
			{97}},\ \bibinfo {pages} {064415} (\bibinfo {year}
		{2018}{\natexlab{a}})}\BibitemShut {NoStop}%
	\bibitem [{\citenamefont {Roscher}\ \emph {et~al.}(2018)\citenamefont
		{Roscher}, \citenamefont {Buessen}, \citenamefont {Scherer}, \citenamefont
		{Trebst},\ and\ \citenamefont {Diehl}}]{Roscher2018}%
	\BibitemOpen
	\bibfield  {author} {\bibinfo {author} {\bibfnamefont {D.}~\bibnamefont
			{Roscher}}, \bibinfo {author} {\bibfnamefont {F.~L.}\ \bibnamefont
			{Buessen}}, \bibinfo {author} {\bibfnamefont {M.~M.}\ \bibnamefont
			{Scherer}}, \bibinfo {author} {\bibfnamefont {S.}~\bibnamefont {Trebst}}, \
		and\ \bibinfo {author} {\bibfnamefont {S.}~\bibnamefont {Diehl}},\ }\bibfield
	{title} {\enquote {\bibinfo {title} {{Functional renormalization group
					approach to $\text{SU}(N)$ Heisenberg models: Momentum-space renormalization
					group for the large-$N$ limit}},}\ }\href {\doibase
		10.1103/PhysRevB.97.064416} {\bibfield  {journal} {\bibinfo  {journal} {Phys.
				Rev. B}\ }\textbf {\bibinfo {volume} {97}},\ \bibinfo {pages} {064416}
		(\bibinfo {year} {2018})}\BibitemShut {NoStop}%
	\bibitem [{\citenamefont {Luttinger}\ and\ \citenamefont
		{Tisza}(1946)}]{Luttinger1946}%
	\BibitemOpen
	\bibfield  {author} {\bibinfo {author} {\bibfnamefont {J.~M.}\ \bibnamefont
			{Luttinger}}\ and\ \bibinfo {author} {\bibfnamefont {L.}~\bibnamefont
			{Tisza}},\ }\bibfield  {title} {\enquote {\bibinfo {title} {{Theory of Dipole
					Interaction in Crystals}},}\ }\href {\doibase 10.1103/PhysRev.70.954}
	{\bibfield  {journal} {\bibinfo  {journal} {Phys. Rev.}\ }\textbf {\bibinfo
			{volume} {70}},\ \bibinfo {pages} {954} (\bibinfo {year} {1946})}\BibitemShut
	{NoStop}%
	\bibitem [{\citenamefont {Luttinger}(1951)}]{Luttinger1951}%
	\BibitemOpen
	\bibfield  {author} {\bibinfo {author} {\bibfnamefont {J.~M.}\ \bibnamefont
			{Luttinger}},\ }\bibfield  {title} {\enquote {\bibinfo {title} {{A Note on
					the Ground State in Antiferromagnetics}},}\ }\href {\doibase
		10.1103/PhysRev.81.1015} {\bibfield  {journal} {\bibinfo  {journal} {Phys.
				Rev.}\ }\textbf {\bibinfo {volume} {81}},\ \bibinfo {pages} {1015} (\bibinfo
		{year} {1951})}\BibitemShut {NoStop}%
	\bibitem [{\citenamefont {Henley}(1987)}]{Henley1987}%
	\BibitemOpen
	\bibfield  {author} {\bibinfo {author} {\bibfnamefont {C.~L.}\ \bibnamefont
			{Henley}},\ }\bibfield  {title} {\enquote {\bibinfo {title} {{Ordering by
					disorder: Ground-state selection in fcc vector antiferromagnets}},}\ }\href
	{\doibase 10.1063/1.338570} {\bibfield  {journal} {\bibinfo  {journal} {J.
				Appl. Phys.}\ }\textbf {\bibinfo {volume} {61}},\ \bibinfo {pages} {3962}
		(\bibinfo {year} {1987})}\BibitemShut {NoStop}%
	\bibitem [{Note2()}]{Note2}%
	\BibitemOpen
	\bibinfo {note} {A similar scenario has recently been discussed in the
		context of the $J_1-J_2$ Heisenberg model on the diamond lattice \cite
		{Bergman2007,Buessen2018b}.}\BibitemShut {Stop}%
	\bibitem [{\citenamefont {Reuther}\ \emph {et~al.}(2011)\citenamefont
		{Reuther}, \citenamefont {Thomale},\ and\ \citenamefont
		{Trebst}}]{Reuther2011c}%
	\BibitemOpen
	\bibfield  {author} {\bibinfo {author} {\bibfnamefont {J.}~\bibnamefont
			{Reuther}}, \bibinfo {author} {\bibfnamefont {R.}~\bibnamefont {Thomale}}, \
		and\ \bibinfo {author} {\bibfnamefont {S.}~\bibnamefont {Trebst}},\
	}\bibfield  {title} {\enquote {\bibinfo {title} {{Finite-temperature phase
					diagram of the Heisenberg-Kitaev model}},}\ }\href {\doibase
		10.1103/PhysRevB.84.100406} {\bibfield  {journal} {\bibinfo  {journal} {Phys.
				Rev. B}\ }\textbf {\bibinfo {volume} {84}},\ \bibinfo {pages} {100406}
		(\bibinfo {year} {2011})}\BibitemShut {NoStop}%
	\bibitem [{\citenamefont {Liu}\ and\ \citenamefont
		{Khaliullin}(2019)}]{Khaliullin2018}%
	\BibitemOpen
	\bibfield  {author} {\bibinfo {author} {\bibfnamefont {H.}~\bibnamefont
			{Liu}}\ and\ \bibinfo {author} {\bibfnamefont {G.}~\bibnamefont
			{Khaliullin}},\ }\bibfield  {title} {\enquote {\bibinfo {title}
			{{Pseudo-Jahn-Teller Effect and Magnetoelastic Coupling in Spin-Orbit Mott
					Insulators}},}\ }\href {\doibase 10.1103/PhysRevLett.122.057203} {\bibfield
		{journal} {\bibinfo  {journal} {Phys. Rev. Lett.}\ }\textbf {\bibinfo
			{volume} {122}},\ \bibinfo {pages} {057203} (\bibinfo {year}
		{2019})}\BibitemShut {NoStop}%
	\bibitem [{\citenamefont {Aczel}\ \emph {et~al.}(2019)\citenamefont {Aczel},
		\citenamefont {Clancy}, \citenamefont {Chen}, \citenamefont {Zhou},
		\citenamefont {{Reig-i-Plessis}}, \citenamefont {MacDougall}, \citenamefont
		{Ruff}, \citenamefont {Upton}, \citenamefont {Islam}, \citenamefont
		{Williams}, \citenamefont {Calder},\ and\ \citenamefont {Yan}}]{Aczel2019}%
	\BibitemOpen
	\bibfield  {author} {\bibinfo {author} {\bibfnamefont {A.~A.}\ \bibnamefont
			{Aczel}}, \bibinfo {author} {\bibfnamefont {J.~P.}\ \bibnamefont {Clancy}},
		\bibinfo {author} {\bibfnamefont {Q.}~\bibnamefont {Chen}}, \bibinfo {author}
		{\bibfnamefont {H.~D.}\ \bibnamefont {Zhou}}, \bibinfo {author}
		{\bibfnamefont {D.}~\bibnamefont {{Reig-i-Plessis}}}, \bibinfo {author}
		{\bibfnamefont {G.~J.}\ \bibnamefont {MacDougall}}, \bibinfo {author}
		{\bibfnamefont {J.~P.~C.}\ \bibnamefont {Ruff}}, \bibinfo {author}
		{\bibfnamefont {M.~H.}\ \bibnamefont {Upton}}, \bibinfo {author}
		{\bibfnamefont {Z.}~\bibnamefont {Islam}}, \bibinfo {author} {\bibfnamefont
			{T.~J.}\ \bibnamefont {Williams}}, \bibinfo {author} {\bibfnamefont
			{S.}~\bibnamefont {Calder}}, \ and\ \bibinfo {author} {\bibfnamefont {J.-Q.}\
			\bibnamefont {Yan}},\ }\bibfield  {title} {\enquote {\bibinfo {title}
			{{Revisiting the Kitaev material candidacy of ${\mathrm{Ir}}^{4+}$ double
					perovskite iridates}},}\ }\href {\doibase 10.1103/PhysRevB.99.134417}
	{\bibfield  {journal} {\bibinfo  {journal} {Phys. Rev. B}\ }\textbf {\bibinfo
			{volume} {99}},\ \bibinfo {pages} {134417} (\bibinfo {year}
		{2019})}\BibitemShut {NoStop}%
	\bibitem [{\citenamefont {Khan}\ \emph {et~al.}(2019)\citenamefont {Khan},
		\citenamefont {Prishchenko}, \citenamefont {Skourski}, \citenamefont
		{Mazurenko},\ and\ \citenamefont {Tsirlin}}]{Khan2019}%
	\BibitemOpen
	\bibfield  {author} {\bibinfo {author} {\bibfnamefont {N.}~\bibnamefont
			{Khan}}, \bibinfo {author} {\bibfnamefont {D.}~\bibnamefont {Prishchenko}},
		\bibinfo {author} {\bibfnamefont {Y.}~\bibnamefont {Skourski}}, \bibinfo
		{author} {\bibfnamefont {V.~G.}\ \bibnamefont {Mazurenko}}, \ and\ \bibinfo
		{author} {\bibfnamefont {A.~A.}\ \bibnamefont {Tsirlin}},\ }\bibfield
	{title} {\enquote {\bibinfo {title} {{Cubic symmetry and magnetic frustration
					on the fcc spin lattice in ${\mathrm{K}}_{2}{\mathrm{IrCl}}_{6}$}},}\ }\href
	{\doibase 10.1103/PhysRevB.99.144425} {\bibfield  {journal} {\bibinfo
			{journal} {Phys. Rev. B}\ }\textbf {\bibinfo {volume} {99}},\ \bibinfo
		{pages} {144425} (\bibinfo {year} {2019})}\BibitemShut {NoStop}%
	\bibitem [{\citenamefont {Bain}\ and\ \citenamefont {Berry}(2008)}]{Bain2008}%
	\BibitemOpen
	\bibfield  {author} {\bibinfo {author} {\bibfnamefont {G.~A.}\ \bibnamefont
			{Bain}}\ and\ \bibinfo {author} {\bibfnamefont {J.~F.}\ \bibnamefont
			{Berry}},\ }\bibfield  {title} {\enquote {\bibinfo {title} {{Diamagnetic
					Corrections and Pascal's Constants}},}\ }\href {\doibase 10.1021/ed085p532}
	{\bibfield  {journal} {\bibinfo  {journal} {J. Chem. Educ.}\ }\textbf
		{\bibinfo {volume} {85}},\ \bibinfo {pages} {532} (\bibinfo {year}
		{2008})}\BibitemShut {NoStop}%
	\bibitem [{\citenamefont {Bergman}\ \emph {et~al.}(2007)\citenamefont
		{Bergman}, \citenamefont {Alicea}, \citenamefont {Gull}, \citenamefont
		{Trebst},\ and\ \citenamefont {Balents}}]{Bergman2007}%
	\BibitemOpen
	\bibfield  {author} {\bibinfo {author} {\bibfnamefont {D.}~\bibnamefont
			{Bergman}}, \bibinfo {author} {\bibfnamefont {J.}~\bibnamefont {Alicea}},
		\bibinfo {author} {\bibfnamefont {E.}~\bibnamefont {Gull}}, \bibinfo {author}
		{\bibfnamefont {S.}~\bibnamefont {Trebst}}, \ and\ \bibinfo {author}
		{\bibfnamefont {L.}~\bibnamefont {Balents}},\ }\bibfield  {title} {\enquote
		{\bibinfo {title} {Order-by-disorder and spiral spin-liquid in frustrated
				diamond-lattice antiferromagnets},}\ }\href
	{http://dx.doi.org/10.1038/nphys622} {\bibfield  {journal} {\bibinfo
			{journal} {Nat. Phys.}\ }\textbf {\bibinfo {volume} {3}},\ \bibinfo {pages}
		{487} (\bibinfo {year} {2007})}\BibitemShut {NoStop}%
	\bibitem [{\citenamefont {Buessen}\ \emph
		{et~al.}(2018{\natexlab{b}})\citenamefont {Buessen}, \citenamefont {Hering},
		\citenamefont {Reuther},\ and\ \citenamefont {Trebst}}]{Buessen2018b}%
	\BibitemOpen
	\bibfield  {author} {\bibinfo {author} {\bibfnamefont {F.~L.}\ \bibnamefont
			{Buessen}}, \bibinfo {author} {\bibfnamefont {M.}~\bibnamefont {Hering}},
		\bibinfo {author} {\bibfnamefont {J.}~\bibnamefont {Reuther}}, \ and\
		\bibinfo {author} {\bibfnamefont {S.}~\bibnamefont {Trebst}},\ }\bibfield
	{title} {\enquote {\bibinfo {title} {{Quantum Spin Liquids in Frustrated
					Spin-1 Diamond Antiferromagnets}},}\ }\href {\doibase
		10.1103/PhysRevLett.120.057201} {\bibfield  {journal} {\bibinfo  {journal}
			{Phys. Rev. Lett.}\ }\textbf {\bibinfo {volume} {120}},\ \bibinfo {pages}
		{057201} (\bibinfo {year} {2018}{\natexlab{b}})}\BibitemShut {NoStop}%
	\bibitem [{\citenamefont {Elwell}\ and\ \citenamefont
		{Scheel}(1975)}]{Elwell75}%
	\BibitemOpen
	\bibfield  {author} {\bibinfo {author} {\bibfnamefont {D.}~\bibnamefont
			{Elwell}}\ and\ \bibinfo {author} {\bibfnamefont {H.J.}\ \bibnamefont
			{Scheel}},\ }\href@noop {} {\emph {\bibinfo {title} {{Crystal Growth from
					High-Temperature Solutions}}}}\ (\bibinfo  {publisher} {Academic Press,
		London, New York, San Francisco},\ \bibinfo {year} {1975})\BibitemShut
	{NoStop}%
	\bibitem [{\citenamefont {Yuan}\ \emph {et~al.}(2017)\citenamefont {Yuan},
		\citenamefont {Clancy}, \citenamefont {Cook}, \citenamefont {Thompson},
		\citenamefont {Greedan}, \citenamefont {Cao}, \citenamefont {Jeon},
		\citenamefont {Noh}, \citenamefont {Upton}, \citenamefont {Casa},
		\citenamefont {Gog}, \citenamefont {Paramekanti},\ and\ \citenamefont
		{Kim}}]{YJKim2017}%
	\BibitemOpen
	\bibfield  {author} {\bibinfo {author} {\bibfnamefont {B.}~\bibnamefont
			{Yuan}}, \bibinfo {author} {\bibfnamefont {J.~P.}\ \bibnamefont {Clancy}},
		\bibinfo {author} {\bibfnamefont {A.~M.}\ \bibnamefont {Cook}}, \bibinfo
		{author} {\bibfnamefont {C.~M.}\ \bibnamefont {Thompson}}, \bibinfo {author}
		{\bibfnamefont {J.}~\bibnamefont {Greedan}}, \bibinfo {author} {\bibfnamefont
			{G.}~\bibnamefont {Cao}}, \bibinfo {author} {\bibfnamefont {B.~C.}\
			\bibnamefont {Jeon}}, \bibinfo {author} {\bibfnamefont {T.~W.}\ \bibnamefont
			{Noh}}, \bibinfo {author} {\bibfnamefont {M.~H.}\ \bibnamefont {Upton}},
		\bibinfo {author} {\bibfnamefont {D.}~\bibnamefont {Casa}}, \bibinfo {author}
		{\bibfnamefont {T.}~\bibnamefont {Gog}}, \bibinfo {author} {\bibfnamefont
			{A.}~\bibnamefont {Paramekanti}}, \ and\ \bibinfo {author} {\bibfnamefont
			{Y.-J.}\ \bibnamefont {Kim}},\ }\bibfield  {title} {\enquote {\bibinfo
			{title} {{Determination of Hund's coupling in $5d$ oxides using resonant
					inelastic x-ray scattering}},}\ }\href {\doibase 10.1103/PhysRevB.95.235114}
	{\bibfield  {journal} {\bibinfo  {journal} {Phys. Rev. B}\ }\textbf {\bibinfo
			{volume} {95}},\ \bibinfo {pages} {235114} (\bibinfo {year}
		{2017})}\BibitemShut {NoStop}%
	\bibitem [{\citenamefont {Nag}\ \emph {et~al.}(2018)\citenamefont {Nag},
		\citenamefont {Bhowal}, \citenamefont {Chakraborty}, \citenamefont
		{Moretti~Sala}, \citenamefont {Efimenko}, \citenamefont {Bert}, \citenamefont
		{Biswas}, \citenamefont {Hillier}, \citenamefont {Itoh}, \citenamefont
		{Kaushik}, \citenamefont {Siruguri}, \citenamefont {Meneghini}, \citenamefont
		{Dasgupta},\ and\ \citenamefont {Ray}}]{Nag_PRB2018}%
	\BibitemOpen
	\bibfield  {author} {\bibinfo {author} {\bibfnamefont {A.}~\bibnamefont
			{Nag}}, \bibinfo {author} {\bibfnamefont {S.}~\bibnamefont {Bhowal}},
		\bibinfo {author} {\bibfnamefont {A.}~\bibnamefont {Chakraborty}}, \bibinfo
		{author} {\bibfnamefont {M.}~\bibnamefont {Moretti~Sala}}, \bibinfo {author}
		{\bibfnamefont {A.}~\bibnamefont {Efimenko}}, \bibinfo {author}
		{\bibfnamefont {F.}~\bibnamefont {Bert}}, \bibinfo {author} {\bibfnamefont
			{P.~K.}\ \bibnamefont {Biswas}}, \bibinfo {author} {\bibfnamefont {A.~D.}\
			\bibnamefont {Hillier}}, \bibinfo {author} {\bibfnamefont {M.}~\bibnamefont
			{Itoh}}, \bibinfo {author} {\bibfnamefont {S.~D.}\ \bibnamefont {Kaushik}},
		\bibinfo {author} {\bibfnamefont {V.}~\bibnamefont {Siruguri}}, \bibinfo
		{author} {\bibfnamefont {C.}~\bibnamefont {Meneghini}}, \bibinfo {author}
		{\bibfnamefont {I.}~\bibnamefont {Dasgupta}}, \ and\ \bibinfo {author}
		{\bibfnamefont {S.}~\bibnamefont {Ray}},\ }\bibfield  {title} {\enquote
		{\bibinfo {title} {{Origin of magnetic moments and presence of spin-orbit
					singlets in ${\mathrm{Ba}}_{2}{\mathrm{YIrO}}_{6}$}},}\ }\href {\doibase
		10.1103/PhysRevB.98.014431} {\bibfield  {journal} {\bibinfo  {journal} {Phys.
				Rev. B}\ }\textbf {\bibinfo {volume} {98}},\ \bibinfo {pages} {014431}
		(\bibinfo {year} {2018})}\BibitemShut {NoStop}%
	\bibitem [{\citenamefont {Bl{\"{o}}chl}(1994)}]{Blochl1994}%
	\BibitemOpen
	\bibfield  {author} {\bibinfo {author} {\bibfnamefont {P.~E.}\ \bibnamefont
			{Bl{\"{o}}chl}},\ }\bibfield  {title} {\enquote {\bibinfo {title} {{Projector
					augmented-wave method}},}\ }\href
	{http://prb.aps.org/abstract/PRB/v50/i24/p17953{\_}1} {\bibfield  {journal}
		{\bibinfo  {journal} {Phys. Rev. B}\ }\textbf {\bibinfo {volume} {50}},\
		\bibinfo {pages} {17953} (\bibinfo {year} {1994})}\BibitemShut {NoStop}%
	\bibitem [{\citenamefont {Kresse}\ and\ \citenamefont
		{Furthm{\"{u}}ller}(1996)}]{Kresse1996}%
	\BibitemOpen
	\bibfield  {author} {\bibinfo {author} {\bibfnamefont {G.}~\bibnamefont
			{Kresse}}\ and\ \bibinfo {author} {\bibfnamefont {J.}~\bibnamefont
			{Furthm{\"{u}}ller}},\ }\bibfield  {title} {\enquote {\bibinfo {title}
			{{Efficient iterative schemes for ab initio total-energy calculations using a
					plane-wave basis set}},}\ }\href {http://www.ncbi.nlm.nih.gov/pubmed/9984901}
	{\bibfield  {journal} {\bibinfo  {journal} {Phys. Rev. B}\ }\textbf {\bibinfo
			{volume} {54}},\ \bibinfo {pages} {11169} (\bibinfo {year}
		{1996})}\BibitemShut {NoStop}%
	\bibitem [{\citenamefont {Perdew}\ \emph {et~al.}(1996)\citenamefont {Perdew},
		\citenamefont {Burke},\ and\ \citenamefont {Ernzerhof}}]{Perdew1996}%
	\BibitemOpen
	\bibfield  {author} {\bibinfo {author} {\bibfnamefont {J.~P.}\ \bibnamefont
			{Perdew}}, \bibinfo {author} {\bibfnamefont {K.}~\bibnamefont {Burke}}, \
		and\ \bibinfo {author} {\bibfnamefont {M.}~\bibnamefont {Ernzerhof}},\
	}\bibfield  {title} {\enquote {\bibinfo {title} {Generalized gradient
				approximation made simple},}\ }\href {\doibase 10.1103/PhysRevLett.77.3865}
	{\bibfield  {journal} {\bibinfo  {journal} {Phys. Rev. Lett.}\ }\textbf
		{\bibinfo {volume} {77}},\ \bibinfo {pages} {3865} (\bibinfo {year}
		{1996})}\BibitemShut {NoStop}%
	\bibitem [{\citenamefont {Dudarev}\ \emph {et~al.}(1998)\citenamefont
		{Dudarev}, \citenamefont {Savrasov}, \citenamefont {Humphreys},\ and\
		\citenamefont {Sutton}}]{Dudarev1998}%
	\BibitemOpen
	\bibfield  {author} {\bibinfo {author} {\bibfnamefont {S.~L.}\ \bibnamefont
			{Dudarev}}, \bibinfo {author} {\bibfnamefont {S.~Y.}\ \bibnamefont
			{Savrasov}}, \bibinfo {author} {\bibfnamefont {C.~J.}\ \bibnamefont
			{Humphreys}}, \ and\ \bibinfo {author} {\bibfnamefont {A.~P.}\ \bibnamefont
			{Sutton}},\ }\bibfield  {title} {\enquote {\bibinfo {title}
			{{Electron-energy-loss spectra and the structural stability of nickel oxide:
					An LSDA+U study}},}\ }\href {\doibase 10.1103/PhysRevB.57.1505} {\bibfield
		{journal} {\bibinfo  {journal} {Phys. Rev. B}\ }\textbf {\bibinfo {volume}
			{57}},\ \bibinfo {pages} {1505} (\bibinfo {year} {1998})}\BibitemShut
	{NoStop}%
\end{thebibliography}
%

\appendix

\renewcommand{\theequation}{A\arabic{equation}}
\renewcommand{\thefigure}{A\arabic{figure}}
\renewcommand{\thetable}{A\,\Roman{table}}
\setcounter{figure}{0}

\section{Single-crystal growth and characterization}

We have grown single crystals of \BaCeIrO\ by melt solution growth 
using BaCl$_2$ as melt solvent and BaCO$_3$ (Merck, p.a.), IrO$_2$ (Chempur, 99.9\,\%) and 
CeO$_2$ (Auer Remy, 99.9\,\%) as educts for the crystals. 
Due to the moderate solubility of metal oxides in halide melts \cite{Elwell75}, a ratio flux/crystal of 15/1 
was used to achieve sufficient dissolution of the oxides. 
The crucible was sealed with a lid to prevent evaporation of BaCl$_2$ from the melt solution. 
Within three weeks single crystals of about 1\,mm$^3$ size were obtained. 
The black single crystals were separated from the flux by dissolving the flux in deionized water 
and analyzed by energy-dispersive x-ray scattering. 

\begin{figure}[b]
	\centering
	\includegraphics[width=0.49\columnwidth]{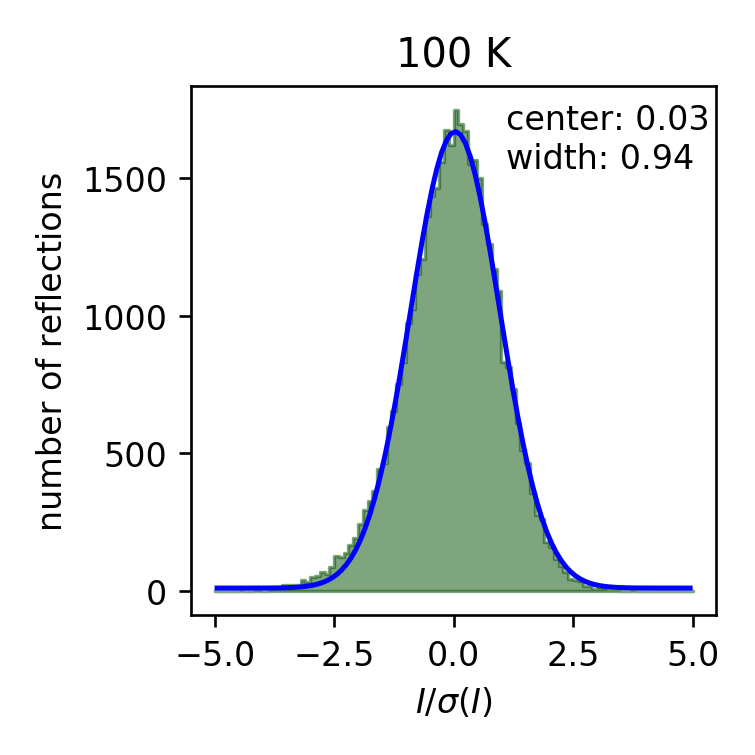}
	\includegraphics[width=0.49\columnwidth]{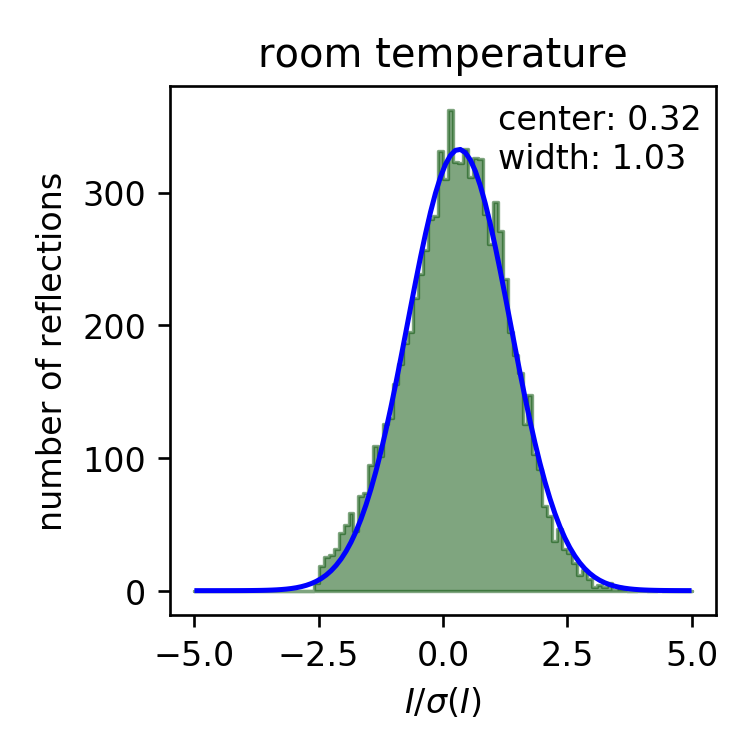}
	\caption{Results of single-crystal x-ray diffraction. 
		Histogram of the intensities of peaks forbidden in $Fm\bar{3}m$ divided by their respective error bars. Left: $T$\,=\,100\,K.\@ Right: room temperature. Blue lines: Gaussian fits. 
		Note that negative values of the intensities arise due to the subtraction of a small overall background.
		The data do not yield any significant evidence for reflections of a non-cubic structure. 
	}
	\label{fig:XRD}
\end{figure}

Our single-crystal x-ray diffraction data strongly support a cubic structure 
	of Ba$_2$CeIrO$_6$, as explained in the main text. 	
	Figure \ref{fig:XRD} shows the distribution of the observed peak intensities $I$ divided by their 
	error bars $\sigma(I)$ for all reflections that are not allowed in space group $Fm\bar{3}m$. 
	The width of the distributions is approximately 1, which is an indication for meaningful 
	statistical errors of the intensities.
	Both at room temperature and at 100\,K, we find a Gaussian profile peaking at a value of 
	$I/\sigma(I)$ close to zero or much smaller than 1, i.e., the intensities of forbidden peaks 
	are negligible within the experimental error bars.

The excellent agreement of the RIXS spectra shown in Fig.\ 3 with the expectations for the 
	spin-orbit exciton provides an unambiguous fingerprint of the Ir$^{4+}$ valence state. 
	RIXS spectra for $5d^4$ Ir$^{5+}$ are distinctly different, as reported for, e.g., the double perovskites 
	Sr$_2$YIrO$_6$ and Ba$_2$YIrO$_6$ \cite{YJKim2017,Nag_PRB2018}. In \BaCeIrO, the Ir$^{4+}$ valence means 
	that also the Ce ions are tetravalent, as claimed before \cite{Wakeshima2000} based on the dependence 
of the lattice parameters on the ionic radius of the lanthanide ions $Ln$ in Ba$_2$$Ln$IrO$_6$. 

\begin{figure}[t]
	\centering
	\includegraphics[width=0.85\columnwidth]{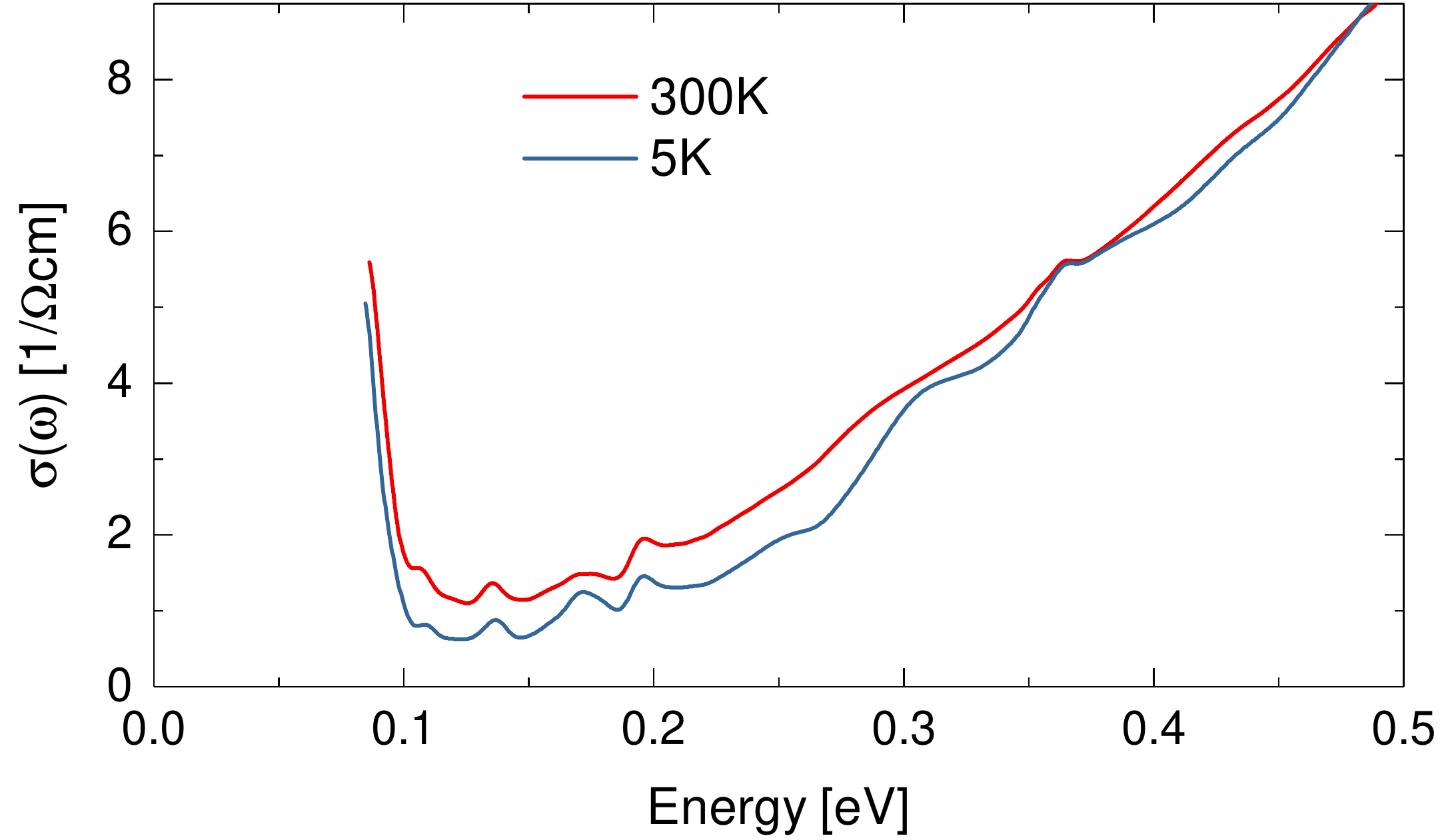}
	\caption{Optical conductivity of \BaCeIrO, showing the onset of excitations across the Mott gap. }
	\label{fig:optics}
\end{figure}

The Mott-insulating character of \BaCeIrO\ is demonstrated by the very low value of the optical conductivity 
in the mid-infrared range, $\sigma_1(\omega)\! \approx \! 1$\,($\Omega$cm)$^{-1}$, see Fig.\ \ref{fig:optics}. 
Using a Bruker IFS 66/v Fourier-transform infrared spectrometer, we measured the transmittance 
on a single crystal with a thickness of ($30\pm 5$)\,$\mu$m. 
On the low-frequency side, the accessible frequency range is cut off by strong phonon absorption suppressing 
the transmittance. 
The steep edge in $\sigma_1(\omega)$ at about 0.1\,eV corresponds to the upper limit for single phonon 
absorption, while the weak features up to about 0.2\,eV can be attributed to multi-phonon absorption. 
On the high-frequency side, the transmittance is suppressed by electron-hole excitations across the gap, 
giving rise to the increase of $\sigma_1(\omega)$ above about 0.2\,eV.\@

\section{Derivation of microscopic model}

\noindent{\em Ab initio calculations.-- }
In order to calculate Heisenberg-type and Kitaev-type exchange constants, we used the 
projector augmented-wave (PAW) method \cite{Blochl1994} as realized in the pseudopotential VASP code \cite{Kresse1996}. 
The exchange-correlation potential was chosen in the form proposed by Perdew, Burke, and Ernzerhof \cite{Perdew1996}.
Electronic correlations and spin-orbit coupling were considered in the framework of the GGA+U+SOC 
formalism \cite{Dudarev1998} with $U\!-\!J_H$\,=\,$2.0$\,eV \cite{Kim2008}, where $U$ and $J_H$ 
denote the on-site Coulomb repulsion and intra-atomic Hund's exchange, respectively. 
The integration was performed on a $7 \times 7 \times 7$ mesh of the Brillouin zone. 
We calculated total energies of three magnetic configurations (FM, AFM type I and type II) to extract
nearest and next-nearest neighbor exchange constants $J_1$ and $J_2$. 
The Kitaev coupling $K$ was computed via the difference of the total energies of 
two AFM type I configurations with spins pointing along $a$ and $c$, respectively.

\vskip 3mm

\noindent{\em Perturbative approach.-- }
For the perturbative calculation of exchange couplings, we use one-hole and two-hole eigenstates and energies obtained by numerical diagonalization of the single-site 
Hamiltonian
\bea 
\! H_{\rm site} \! &=&\! -i \frac{\lambda}{2} \sum_{\ell m n} \epsilon_{\ell m n} d^\dg_{\ell\alpha} d^\pdg_{m\beta} \sigma^n_{\alpha\beta} \! + \!
\frac{U}{2} n_{\rm tot}^2
\! -\!  \frac{5 J_H}{2} \sum_{\ell<\ell'} n^\pdg_{\ell}
n^\pdg_{\ell'} \nonumber \\
\! &-&\! 2 J_H \sum_{\ell< \ell'} \vec S^\pdg_{\ell}
\cdot \vec S^\pdg_{\ell'} + J_H \sum_{\ell \neq \ell'} d^\dg_{\ell\upa}
d^\dg_{\ell\dna} d^\pdg_{\ell' \dna} d^\pdg_{\ell' \upa},
\label{Hint}
\eea
with $\lambda$\,=\,$0.43$\,eV deduced from our RIXS data, Hund's coupling $J_H$\,=\,$0.25$\,eV \cite{YJKim2017}
and $U$\,=\,$2.25$\,eV as a typical estimate for the Hubbard repulsion (which leads to $U\!-\!J_H$\,=\,$2.0$\,eV 
as in the {\em ab initio} calculations). For the inter-site Hamiltonian, we use
$H_{\rm hop}$\,=\,$\sum_{\la i,j\ra,\sigma} {\bf d}^\dg_{i\sigma}\cdot{\rm\bf T}^\pdg_{ij}\cdot{\bf d}^\pdg_{j\sigma}$, 
with ${\bf d}^\dg$ representing the creation operators for the three $t_{2g}$ orbitals, 
$(d^\dg_{yz},d^\dg_{xz},d^\dg_{xy})$. 
For the hopping amplitudes $t_{\alpha-\beta}$ between a pair of orbitals ($\alpha, \beta$) on 
nearest-neighbor sites $(i, i+\hat{x}+\hat{y})$, we retain only the by far dominant one $t_{xy-xy} \! \approx \! -150$\,meV, 
which is an order of magnitude larger than $t_{xz-yz}$ which again is larger than $t_{xz-xz}$\,=\,$t_{yz-yz}$,
while others vanish by symmetry. 
For second-neighbor pairs $(i,i+2\hat{x})$, we employ 
$t_{xy-xy}$\,=\,$t_{xz-xz} \! \approx \! 30$\,meV,
while other hopping amplitudes vanish by symmetry.
For the cubic {\em fcc} lattice, the corresponding hopping amplitudes between all other nearest-neighbor 
or next-nearest neighbor pairs are determined by symmetry. 
To extract the two-site exchange Hamiltonian, we carry out second-order degenerate perturbation theory in $t/U$, evaluating matrix elements numerically using exact single-site eigenfunctions and energies. 
The resulting exchange Hamiltonian has dominant nearest-neighbor Heisenberg exchange interaction $J_1$ 
with subdominant Kitaev and second-neighbor Heisenberg terms as listed in the main text 
and a negligible $\Gamma \lesssim 0.05 J_1$ exchange term.

\section{Antiferromagnetic character of Kitaev coupling}

The $j$\,=\,$1/2$ wave function is given by 
\begin{equation}
|\frac{1}{2},{\rm up}\rangle = \frac{1}{\sqrt{3}}\left( |yz,\downarrow\rangle + i |xz,\downarrow\rangle 
+ |xy,\uparrow\rangle\right) \, .
\label{eq:j12}
\end{equation}
We address superexchange interactions between two sites A and B in the $xy$ plane. 
For comparison, we first consider edge-sharing geometry with $90^\circ$ Ir-O-Ir bonds \cite{Jackeli2009} 
as approximately realized in the honeycomb iridates. In this case, the hopping $t_{xy-xy}^{90^\circ}$ between 
$xy_A$ and $xy_B$ via an intermediate O ligand vanishes by symmetry. 
There are two finite hopping contributions, between $xz_A$ and $yz_B$ and between $yz_A$ and $xz_B$. 
The Heisenberg interaction vanishes due to destructive interference between these two. 
This destructive interference originates from the phase factor $i$ in the $j$\,=\,$1/2$ wavefunction, 
see Eq.\ \eqref{eq:j12}. In contrast, Kitaev exchange remains finite and is ferromagnetic. 
The ferromagnetic character arises from the virtual intermediate state with two holes on the same site 
occupying the $xz$ and $yz$ orbitals. The energy of this intermediate state is lower for parallel spins 
in $xz$ and $yz$, which corresponds to parallel $j$\,=\,$1/2$ pseudo-spins, see Eq.\ \eqref{eq:j12}. 

The same ferromagnetic contribution to Kitaev exchange is present on the $fcc$ lattice as well 
where the Ir-O-Ir hopping of edge-sharing geometry has to be replaced by Ir-O-O-Ir hopping. 
However, the size of this ferromagnetic contribution to Kitaev exchange is negligibly small on the $fcc$ lattice 
due to the very small hopping between $xz$ and $yz$ orbitals. 
In contrast to the edge-sharing geometry discussed above, the hopping $xy_A \rightarrow xy_B$  via the 
two intermediate O ligands does not vanish on the $fcc$ lattice, it is in fact   the dominant hopping term 
within the $xy$ plane. 
For parallel pseudo-spins $|\frac{1}{2},{\rm up}\rangle_A \, |\frac{1}{2},{\rm up}\rangle_B$, 
the configuration $|xy,\uparrow\rangle_A \, |yz,\downarrow\rangle_B$ in orbital/spin basis carries finite weight, 
see Eq.\ \eqref{eq:j12}. 
Hopping $xy_A \rightarrow xy_B$ then yields the doubly occupied virtual state 
$|xy,\uparrow\rangle_B \, |yz,\downarrow\rangle_B$, and a second hopping process $xy_B \rightarrow xy_A$
brings us back to the ground state without any spin flip. This exchange will lead to an Ising-like $S_A^z S_B^z$ term. Equivalent to the case discussed above, the lowest energy in the virtual doubly occupied state is realized 
for two parallel spins, $|xy,\uparrow\rangle_B \, |yz,\uparrow\rangle_B$, 
which for this combination of orbitals corresponds to \textit{antiparallel} pseudo-spins, see Eq.\ \eqref{eq:j12}. 
As a result, pure $xy_A \rightarrow xy_B$ hopping within the $xy$ plane in combination with Hund's coupling 
generates a small \textit{antiferromagnetic} Kitaev coupling $K \propto J_1 \cdot J_H/U$.  
By symmetry, it couples the $z$ ($x$, $y$) component of the moments for two nearest-neighbor sites 
within the $xy$ ($yz$, $zx$) plane.

\end{document}